\documentclass[10pt,aps,prb,showpacs,floatfix,floats,superscriptaddress,twocolumn]{revtex4-1}
\usepackage{graphicx,latexsym}
\usepackage{dcolumn}
\usepackage{amssymb,amsmath,bm}
\pdfoutput=1

\begin{document}

%-----------------------------------------------------------------
\title{Cavity-photon contribution to the effective interaction\\ of electrons
       in parallel quantum dots}

\author{Vidar Gudmundsson}
\email{vidar@hi.is}
\affiliation{Science Institute, University of Iceland, Dunhaga 3, IS-107 Reykjavik, Iceland}
\author{Anna Sitek}
\affiliation{Science Institute, University of Iceland, Dunhaga 3, IS-107 Reykjavik, Iceland}
\affiliation{Department of Theoretical Physics, Wroc{\l}aw University of Technology, 50-370 Wroc{\l}aw, Poland}
\author{Nzar Rauf Abdullah}
\affiliation{Physics Department, Faculty of Science and Science Education, School of Science, 
             University of Sulaimani, Kurdistan Region, Iraq}
\affiliation{Science Institute, University of Iceland, Dunhaga 3, IS-107 Reykjavik, Iceland}
\author{Chi-Shung Tang}
\email{cstang@nuu.edu.tw}
\affiliation{Department of Mechanical Engineering, National United University, Miaoli 36003, Taiwan}
\author{Andrei Manolescu}
\email{manoles@ru.is}
\affiliation{School of Science and Engineering, Reykjavik University, Menntavegur 
             1, IS-101 Reykjavik, Iceland}

%
%----------------------------------------------------------------

\begin{abstract}
A single cavity photon mode is expected to modify the Coulomb interaction of an
electron system in the cavity. Here we investigate this phenomena in a parallel 
double quantum dot system. We explore properties of the closed system and the 
system after it has been opened up for electron transport. We show how results 
for both cases support the idea that the effective electron-electron interaction 
becomes more repulsive in the presence of a cavity photon field.
This can be understood in terms of the cavity photons dressing the 
polarization terms in the effective mutual electron interaction leading to nontrivial 
delocalization or polarization of the charge in the double parallel dot potential.
In addition, we find that the effective repulsion of the electrons can be reduced by
quadrupolar collective oscillations excited by an external classical 
dipole electric field.
\end{abstract}


\maketitle

\section{Introduction}
Experiments on atom or electronic systems of various types in photonic cavities
have been opening up a new and exciting venue to test and manipulate photon-matter
interactions in the strong coupling 
limit.\cite{PhysRevLett.50.1903,Peterson380:2012,PhysRevB.86.085316,Gallardo:10,Frey11:01,PhysRevB.83.205301}

Modeling of these time-dependent systems has commonly been aimed at evaluating 
their steady state properties, that can include such diverse observables as
their conductance,\cite{PhysRevB.87.195427} the life time,\cite{ates2009:724}
the broadening,\cite{PhysRevB.82.045306} or the energy shift of certain
many-body states or modes of relevance. Here, we would like to draw attention to 
how the experimentally challenging transient time regime can reveal information
about the interaction of the participating constituents, in our case, electrons
and cavity photons.

Many different formalisms have been used to describe transport in many-body systems on
the nanometer scale. Most common have been methods built on various 
Green function approaches.\cite{RevModPhys.58.323,Datta00:253} Less common has been the 
use of master equations,\cite{Harbola06:235309,Timm08:195416} or a generalized master
equation in combination with configuration interaction (CI),\cite{Moldoveanu10:155442}
where the Hamiltonian matrix is diagonalized in a truncated Fock many-body 
space.\cite{Pfannkuche91:13132,Maksym90:108}  
Between these two approaches is an interesting correspondence. When dealing with a system
with two types of interactions, like the electron-electron interaction, and the electron-photon
interaction, one typically may start by building a Green function for the electron including
the mutual electron interaction. A second step would be to evaluate the Green function
of the full system dressed by the photons.\cite{PhysRevA.90.055802} 
Similarly, in a CI-approach, one starts by diagonalizing the Hamiltonian matrix
for the Coulomb interacting electrons, constructs a new many-body Fock-basis from the 
states of the Coulomb interacting electrons and possibly the photon number operator, and diagonalizes
then the full Hamiltonian matrix.\cite{Gudmundsson:2013.305} This approach has been
termed: ``Stepwise introduction of model complexity''. Theoretically, similar structure and accuracy
could be accomplished for the static properties in the two very different formalism, though their
actual potential depends strongly on the system of interest.  
Their difference becomes more obvious for the dynamic or the transport part.    

Here, our goal is to apply the latter strategy to analyze the effective interaction
of electrons in a very special system.
We focus our attention on the interaction of two electrons in a parallel double
dot system. We want to answer the question if it is possible to determine how
the interaction of the two electrons with a single photon mode in the system
modifies their repulsion. How is this effective electron-electron interaction? Does
it depend on how the question is approached, using static or dynamic properties
of the system.
This question naturally arises when the energy spectrum of the Jaynes-Cummings model \cite{Jaynes63:89}
is evaluated for strong electron-photon coupling and photon energy out of
resonance with the two atom levels.\cite{Feranchuk96:4035} In this case a strong negative 
dispersion is found with increasing atom-photon coupling, but the model was not intended for
the off-resonance situation, and it is known that the diamagnetic electron-photon interaction has to be 
included in a related model of electrons and cavity photons in a short
quantum wire.\cite{Jonasson2011:01,PhysRevE.86.046701}  

We explore these question for the case of a parallel double dot system with two 
interacting electrons closed with respect to electron exchange with the environment, 
and compare the results with the situation in which the system is the open version of the model, 
when it is coupled to two external leads with a bias voltage that promotes current through it.
We hope to convince the reader that the geometry of this particular system furnishes it with 
very special properties worth studying.     

The paper is organized as follows: First we introduce the components of the static model,
the many-body formalism for exactly interacting electrons and cavity photons in a 
confined system with a continuous geometry describing two parallel dots embedded in a
short quantum wire. We explore its static and dynamics properties with increasing electron-photon 
interaction. The dynamic properties are obtained by observing the response of the system to
an excitation with an external classical electric dipole pulse of short duration.
We then proceed to open the system up to transport by coupling it to two external 
electron leads and investigate the effects of varying the electron-photon interaction 
on the time-dependent currents to and from the system, the charge in it, and its distribution
through it.   

\section{Closed system}
\begin{figure}[htbq]
      \includegraphics[width=0.44\textwidth,angle=0,viewport=20 20 770 714,clip]{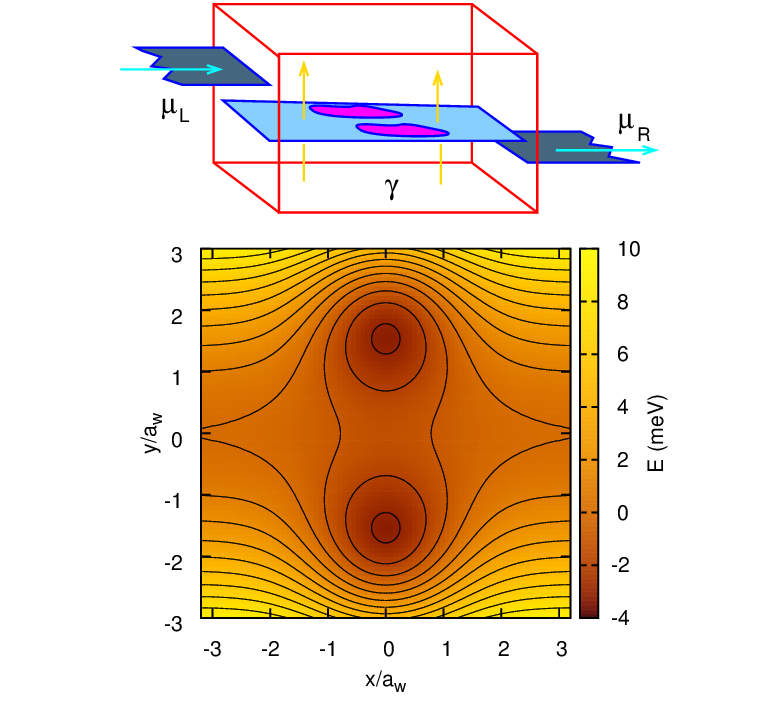}
      \caption{Schema of the leads, the central system with double dots and the 
               cavity (upper panel). The transport direction is indicated by magenta
               arrows, and the photon mode with golden arrows. 
               The potential landscape defining the two parallel
               quantum dots in a short parabolically confined quantum wire (lower panel). 
               The effective magnetic length $a_w=23.8$ nm, $\hbar\Omega_0=2.0$ meV, and $B=0.1$ T.}
      \label{W_2QD}
\end{figure}

In the closed version of the system we consider two electrons in the parallel double-dot
potential landscape shown in Fig.\ \ref{W_2QD} interacting with the Coulomb interaction
that is accounted for via a configuration interaction (CI) approach in the basis of 
noninteracting two-electron states.\cite{Gudmundsson:2013.305} We assume GaAs parameters,
$\kappa =12.4$, and $m^*=0.067m_e$.

The potential landscape of the short quantum wire and the embedded dots is described by
\begin{align}
      V(x,y) =& \left[\frac{1}{2}m^*\Omega_0^2y^2\right.\nonumber\\
             +& \left. V_d\sum_{i=1}^2\exp{\left\{-(\beta x)^2+\beta^2(y-d_i)^2\right\}} \right]\nonumber\\
             \times&\theta\left(\frac{L_x}{2}-x\right)\theta\left(\frac{L_x}{2}+x\right), 
\end{align}
with $\hbar\Omega_0 = 2.0$ meV, $V_d = -6.5$ meV, $\beta = 0.03$ nm$^{-1}$, 
$d_1=-50$ nm, $d_2=-50$ nm, $L_x = 150$ nm, and $\theta$ is the Heaviside 
step function. 

The Hamilton operator for the closed system is
\begin{align}
      H_\mathrm{S} =& \int d^2r \psi^\dagger (\mathbf{r})\left\{\frac{\pi^2}{2m^*}+
        V(\mathbf{r})\right\}\psi (\mathbf{r})
        + H_\mathrm{EM} + H_\mathrm{Coul}\nonumber\\ 
        -&\frac{1}{c}\int d^2r\;\mathbf{j}(\mathbf{r})\cdot\mathbf{A}_\gamma
        -\frac{e}{2m^*c^2}\int d^2r\;\rho(\mathbf{r}) A_\gamma^2,
\label{Hclosed}
\end{align}
with 
\begin{equation}
      {\bm{\pi}}=\left(\mathbf{p}+\frac{e}{c}\mathbf{A}_{\mathrm{ext}}\right),
\end{equation}
where $\mathbf{A}_{\mathrm{ext}}$ is an external vector potential producing a homogeneous
small magnetic field $B=0.1$ T along the $z$-axis included in the calculations in order to 
break spin and orbital degeneracies to guarantee accuracy of the results. We have tested that 
this magnetic field can be made vanishingly small without qualitatively changing our results.
$H_\mathrm{EM}$ is the Hamiltonian for the cavity photon mode with energy $\hbar\omega$,
and the electron-electron interaction represented by $H_\mathrm{Coul}$. The vector potential 
of the cavity photon field is $\mathbf{A}_\gamma$. The two last terms of the Hamiltonian (\ref{Hclosed})
are the para- and the diamagnetic electron-photon interactions, respectively.
The charge and the charge-current densities are 
\begin{equation}
      \rho       = -e\psi^\dagger\psi, \quad
      \mathbf{j} = -\frac{e}{2m^*}\left\{\psi^\dagger\left({\bm{\pi}}\psi\right)
                 +\left({\bm{\pi}}^*\psi^\dagger\right)\psi\right\},
\end{equation}
The natural length scale defined by the parabolic confinement and the external magnetic field 
is $a_w=\sqrt{\hbar /(m^*\Omega_w)}$ with the cyclotron 
frequency $\omega_c=eB/(m^*c)$, and $\Omega_w^2=\omega_c^2+\Omega_0^2$.

We assume a rectangular photon cavity $(x,y,z)\in\{[-a_\mathrm{c}/2,a_\mathrm{c}/2]
\times [-a_\mathrm{c}/2,a_\mathrm{c}/2]\times [-d_\mathrm{c}/2,d_\mathrm{c}/2]\}$ 
with the short quantum wire centered in the $z=0$ plane. In the Coulomb gauge the
polarization of the electric field is chosen parallel to the transport
in the $x$-direction by selecting the TE$_{011}$ mode, or perpendicular to it by selecting the 
TE$_{101}$ mode 
\begin{equation}
      \mathbf{A}_\gamma (\mathbf{r})=\left({\hat{\mathbf{e}}_x\atop \hat{\mathbf{e}}_y}\right)
      {\cal A}\left\{a+a^{\dagger}\right\}
      \left({\cos{\left(\frac{\pi y}{a_\mathrm{c}}\right)}\atop\cos{\left(\frac{\pi x}{a_\mathrm{c}}\right)}} \right)
      \cos{\left(\frac{\pi z}{d_\mathrm{c}}\right)},
\label{Cav-A}
\end{equation}
for the TE$_{011}$ and TE$_{101}$ mode, respectfully. The strength of the vector potential, ${\cal A}$,
determines the coupling constant $g_\mathrm{EM} = e{\cal A}\Omega_wa_w/c$ leaving a dimensionless coupling 
or polarization tensor
\begin{equation}
      g_{ij} = \frac{a_w}{2\hbar}\left\{\langle i|\hat{\mathbf{e}}\cdot\bm{\pi}|j\rangle + \mathrm{h.c.}\right\},  
\end{equation}
where $|i\rangle$ is a single-electron state of the short quantum wire. We are describing an electron
system with many electron states that will generally not be in resonance with the cavity photon mode
so we retain all resonant and antiresonant terms in the photon creation and annihilation operators 
$a^\dagger$ and $a$ (not applying the rotating wave approximation). 
We assume the cavity to be much larger than the electron
system and neglect the spatial variation of the vector potential $\mathbf{A}_\gamma$. 
In a Fock basis for the Coulomb interacting electrons $\{|\mu )\}$ we arrive at the Hamiltonian
\begin{align}
      H_\mathrm{S}&=\sum_\mu |\mu ) E_\mu(\mu| + \hbar\omega a^{\dagger}a \nonumber\\  
      &+ g_{\mathrm{EM}}\sum_{\mu\nu ij}|\mu )
      \langle\mu |{\cal V}^+ d_i^{\dagger}d_j{\cal V}|\nu\rangle (\nu |\;
      g_{ij}\left\{a+a^\dagger\right\} \nonumber\\
      &+g_{\mathrm{EM}}\left(\frac{g_{\mathrm{EM}}}{\hbar\Omega_w}\right)
      \sum_{\mu\nu i}
      |\mu )\langle\mu |{\cal V}^+ d_i^{\dagger}d_i{\cal V}|\nu\rangle (\nu | \nonumber\\
      &\quad\quad\quad\quad \left\{\left( a^\dagger a+\frac{1}{2}\right) +
      \frac{1}{2}\left( aa+a^\dagger a^\dagger\right)\right\},
\label{H-e-VEMV}
\end{align} 
where $d_i$ and $d_i^{\dagger}$ are the electron annihilation and creation operators for the 
single-electron state $|i\rangle$, and the Fock basis for noninteracting electrons $\{|\mu\rangle\}$
is connected to the interacting electron Fock basis by the unitary transform 
$|\mu ) = \sum_{\alpha}{\cal V}_{\mu\alpha}|\alpha\rangle$. The spectrum and states of this
Hamiltonian (\ref{H-e-VEMV}) are found by diagonalization in a Fock space constructed by
a tensor product of the eigenstates of the photon number operator $N_\gamma =a^\dagger a$ and the 
Fock basis of Coulomb interacting electrons $\{|\mu )\}$. 
In this basis without electron-photon interactions each electron state can be found
with a different number of photons, these state are called photon replicas of the electron states.
The electron-photon interaction makes the photon content of an interacting state deviate from
integer numbers for strong coupling. If the deviation is not large the ``photon replica'' concept
can still be used for the interacting system. The ``first photon replica'' of a certain
state is then the tensor product of that state and the eigenstate of the photon number operator
with one photon.   
The diagonalizations for the Coulomb interaction on one hand, and for the electron-photon
paramagnetic- and diamagnetic interactions are done separately  
in truncated bases to optimize the size needed.\cite{Gudmundsson:2013.305,2040-8986-17-1-015201,Jonasson2011:01}

Here, we come to an essential point in our approach: Researchers have found that for the 
Jaynes-Cumming model\cite{Jaynes63:89} in the strong coupling limit it can be difficult
to achieve convergence in a basis constructed with the photon number operator.\cite{0295-5075-96-1-14003}
We do not encounter this problem for our model when the diamagnetic electron-photon interaction is
included (\ref{Hclosed}),\cite{PhysRevE.86.046701} but as the cavity-photon dressed electron-electron
interaction polarizes the charge density we have to be very careful about including enough 
of higher energy electron states in our bases.  

\begin{figure}[htbq]
      \includegraphics[width=0.44\textwidth,angle=0,viewport=20 10 1520 1560,clip]{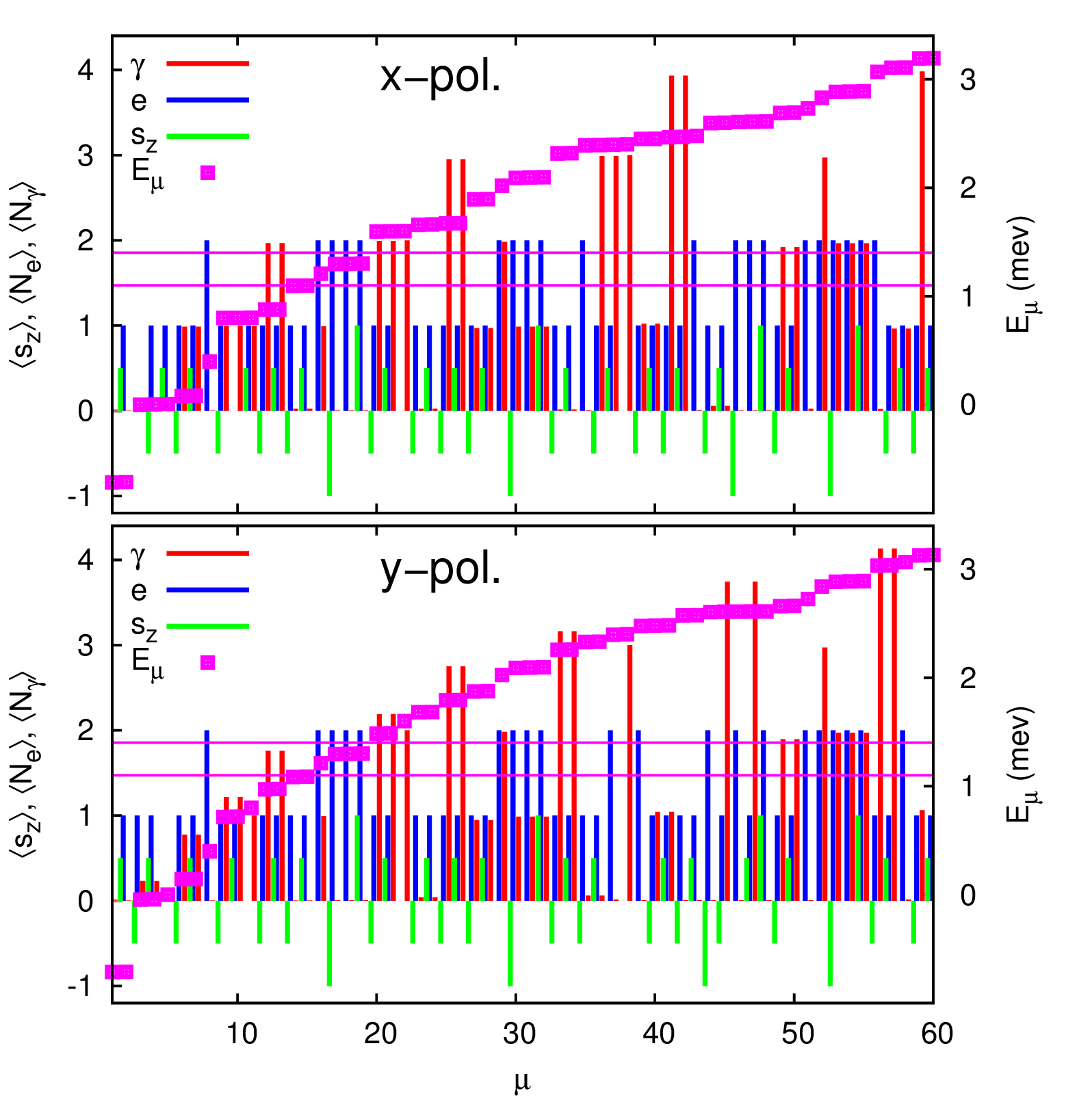}
      \caption{The energy spectra for the 60 lowest many-body states 
               ($E_\mu,\,\mu =1,2,\cdots ,60$) showing their
               electron ($e$), photon ($\gamma$), and spin content ($s_z$). 
               The photon field is $x$-polarized in
               the upper panel, but $y$-polarized in the lower. The left and right chemical
               potentials, $\mu_L$ and $\mu_R$, are shown as horizontal lines for the case 
               of the open system. $\hbar\Omega_0=2.0$ meV, $\hbar\omega =0.8$ meV, 
               $g_\mathrm{EM}=0.1$ meV, $B=0.1$ T, $\kappa =12.4$, and $m^*=0.067m_e$.}
      \label{E-rof}
\end{figure}

The polarization of the photon field will be chosen along the $x$-axis parallel to the electron
transport when the system will be opened, or perpendicular to it, along the $y$-axis.
When calculating matrix elements we assume the wavelength of the cavity photons to be much
larger than the size of the electronic system. 
At the moment the system is closed with respect to electrons and photons.
Since, later we intend to open up the system we include in our calculation states with up to
3 electrons. The energy spectra for the 60 lowest states together with their photon
and spin content are displayed in Fig.\ \ref{E-rof} for the two polarizations of the 
photon field.

The inclusion of the electron-photon interaction causes dispersion of the energy for both one- and two-electron
states with respect to the electron-photon coupling constant $g_\mathrm{EM}$.\cite{Gudmundsson12:1109.4728}
The former can be thought of in terms of an effective potential, and the latter we explain as an effective
cavity-photon dressed electron-electron interaction. 

\begin{figure}[htbq]
      \includegraphics[width=0.44\textwidth,angle=0,viewport=5 10 1440 1020,clip]{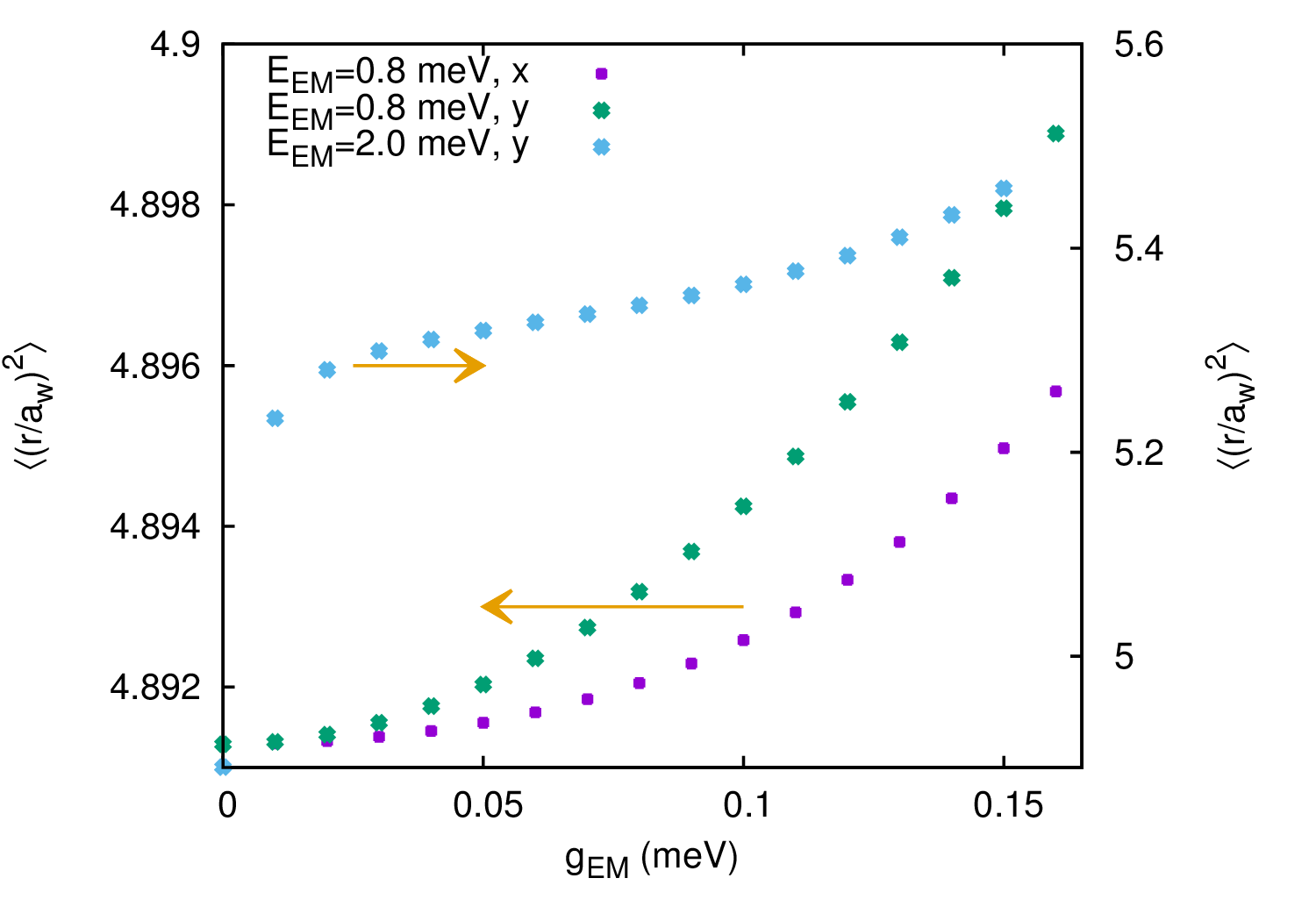}
      \includegraphics[width=0.44\textwidth,angle=0,viewport=5 10 1440 1020,clip]{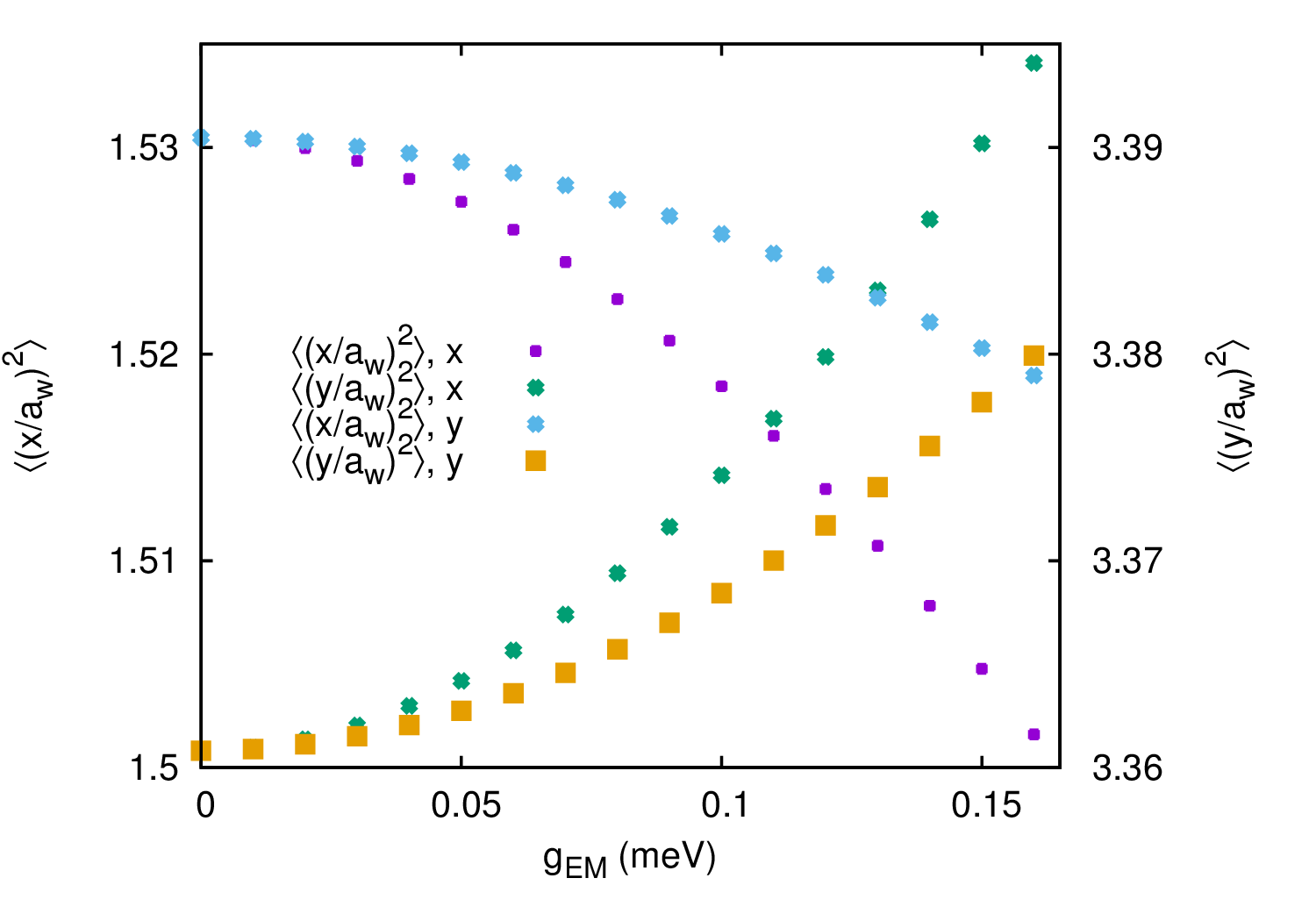}
      \caption{(Upper panel) The expectation value $\langle (r/a_w)^2\rangle$ for the closed system at $t=0$
               for $x$- and $y$-polarization of the photon field. (Lower panel) the expectation values
               $\langle (x/a_w)^2\rangle$ and $\langle (y/a_w)^2\rangle$ for the closed system at $t=0$
               for $x$- and $y$-polarization of the photon field. The photon energy is 
               $E_\mathrm{EM}=\hbar\omega=0.8$ meV except for one curve for comparison in the upper panel.
               $\hbar\Omega_0=2.0$ meV, and $B=0.1$ T.}
      \label{xy-ave}
\end{figure}

We start our study by analyzing the spatial extent of the charge density of the many-body relevant states.
In the upper panel of Fig.\ \ref{xy-ave} we present the average extent of the first photon replica of 
the 2-electron ground state as a function of the coupling $g_\mathrm{EM}$ for the closed system with 
the photon energy $\hbar\omega =0.8$ meV, compared to the case of $\hbar\omega =2.0$ meV when the 
$y$-polarized photon field is in resonance with the confinement energy of the system in the $y$-direction.
The dressing of the electron-electron interaction by the paramagnetic electron-photon term
stretches the charge density in the direction of the polarization of the photon field while 
the dressing by the diamagnetic term spreads it in all directions.
Here, we see that in all cases the extent of the charge density increases, though by a small amount
for the photon field out of resonance. In the lower panel of Fig.\ \ref{xy-ave} we analyze the change 
in the extent of the charge density of the same state, but along the $x$- and $y$-axes. Surprisingly, 
the charge density of the 2-electron state always shrinks in the $x$-direction and expands in the 
$y$-direction with increasing $g_\mathrm{EM}$, irrespective of the polarization of the photon field. 

\begin{figure}[htbq]
      \includegraphics[width=0.48\textwidth,angle=0,viewport=150 20 1200 500,clip]{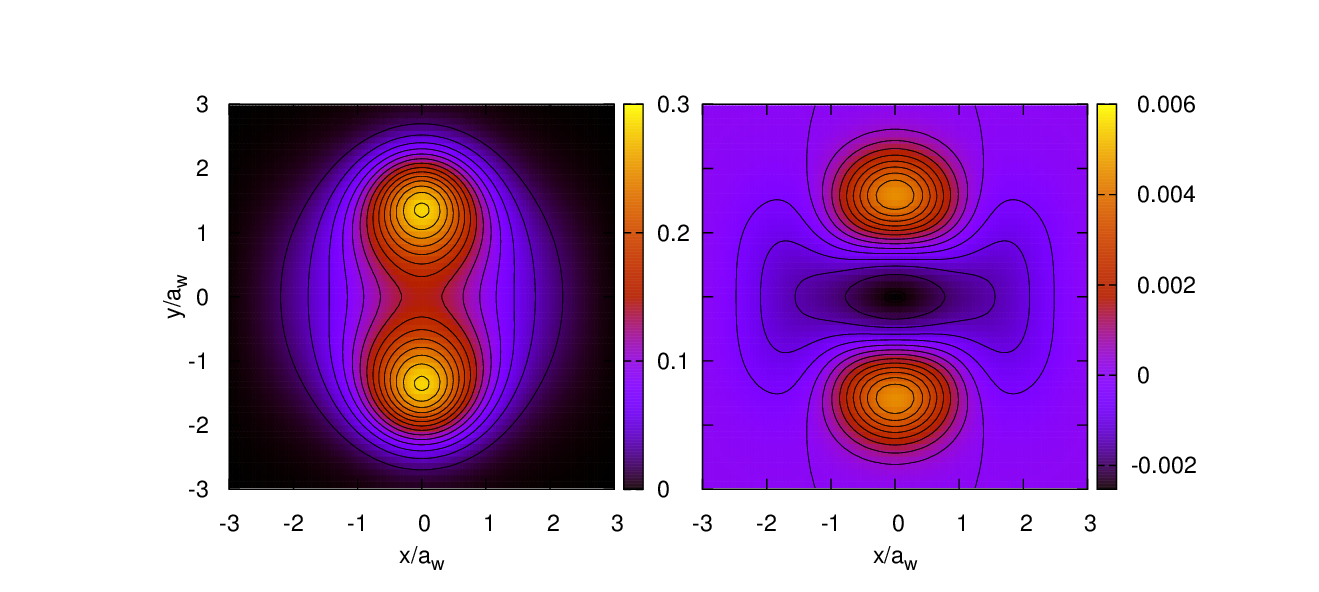}
      \caption{(Left panel) The probability density for ground state for 
               2 electrons and 1 photon with $g_\mathrm{EM}=0$.
               (Right panel) The difference in the charge density 
               $\delta n=n(g_\mathrm{EM}=0.16)-n(0)$ for the same state as in the left panel.
               $x$-polarized photon field, $\hbar\omega=0.8$ meV, $B=0.1$ T.}
      \label{d0}
\end{figure}

To show how this happens for the $x$-polarized photon field we compare in Fig.\ \ref{d0} the probability 
density for the first photon replica of the 2-electron ground state with the changes in the density 
defined as $\delta n=n(g_\mathrm{EM}=0.16)-n(0)$. Clearly, the central section of the density shrinks in
$x$-extent, but together with the stretching in the $y$-direction we notice a stretching in the $x$-direction
for the density located in the quantum dots.

From these observations we conclude that the polarization of the 
charge or probability density of the 2-electron state indicates an enhanced repulsion
between the electrons.

\begin{figure}[htbq]
      \includegraphics[width=0.44\textwidth,angle=0,viewport=170 90 1330 820,clip]{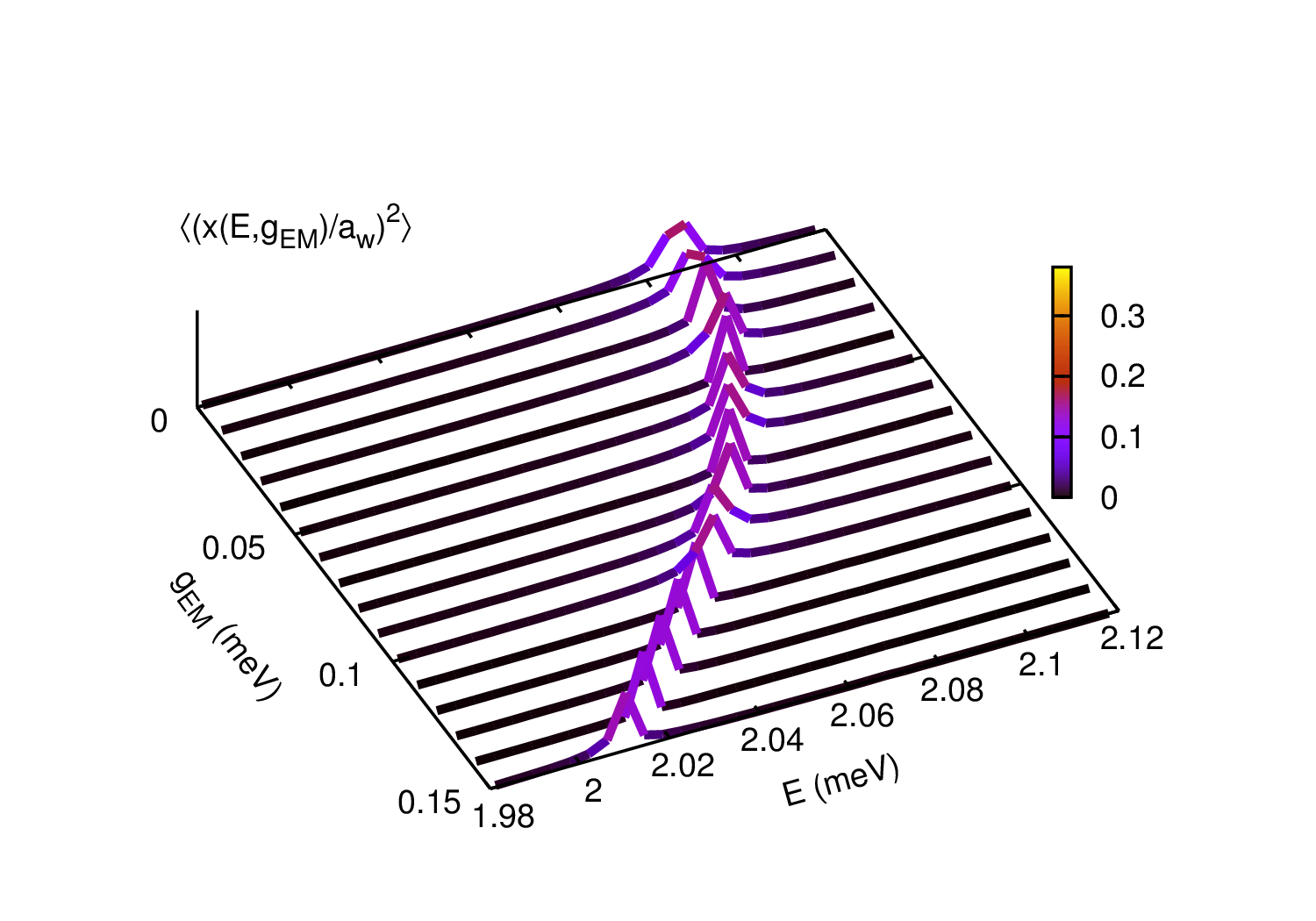}
      \includegraphics[width=0.44\textwidth,angle=0,viewport=170 90 1330 820,clip]{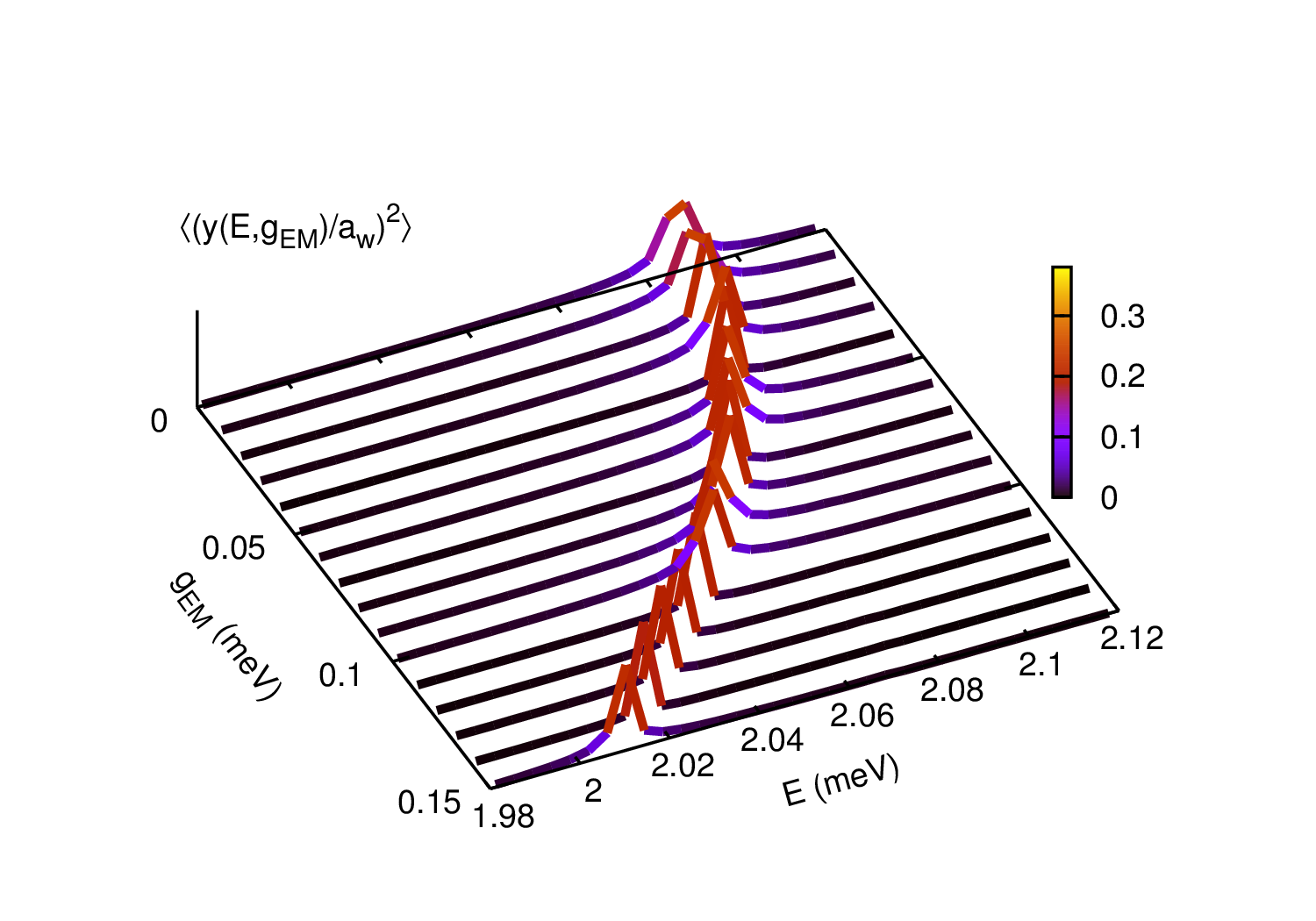}
      \caption{The Fourier transform of the expectation values $\langle (x/a_w)^2\rangle (t)$ (upper), 
               and $\langle (y/a_w)^2\rangle (t)$ (lower) as a function of the electron-photon coupling 
               $g_\mathrm{EM}$. $y$-polarized photon field and excitation pulse. $\hbar\Omega_0=2.0$ meV, and $B=0.1$ T.} 
      \label{Absorption}
\end{figure}

\begin{figure}[htbq]
      \includegraphics[width=0.48\textwidth,angle=0,viewport=10 50 1720 1225,clip]{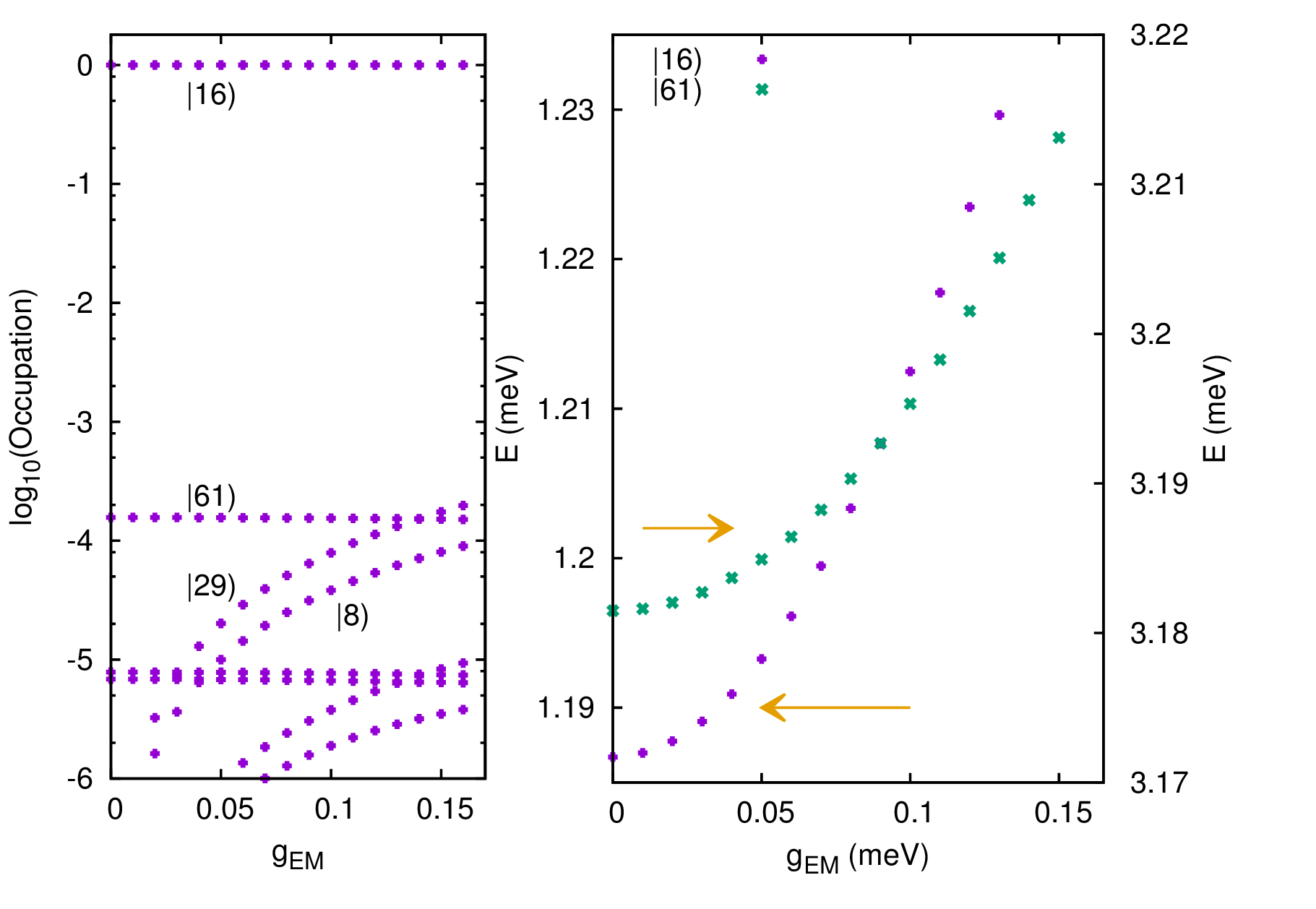}
      \caption{(Left) In the closed system (after the excitation pulse vanishes) 
               the occupation of the states as a function of the electron-photon coupling
               $g_\mathrm{EM}$. (Right) The dispersion of the initial state $|\breve{16})$ and 
               the excited state $|\breve{61})$ (both with 2 electrons and 1 photon). 
               $y$-polarization of the photon field and dipole excitation pulse in the same
               direction. States $|\breve{8})$ and $|\breve{29})$ are 2-electron states with
               0 and 2 photons, respectively. $E_\mathrm{EM}=\hbar\omega=0.8$ meV,
               $\hbar\Omega_0=2.0$ meV, and $B=0.1$ T. The arrows indicate the respective 
               equidistant $y$-axis.}
      \label{E-spectrum}
\end{figure}

\begin{figure}[htbq]
      \includegraphics[width=0.30\textwidth,angle=0,viewport=270 40 1260 890,clip]{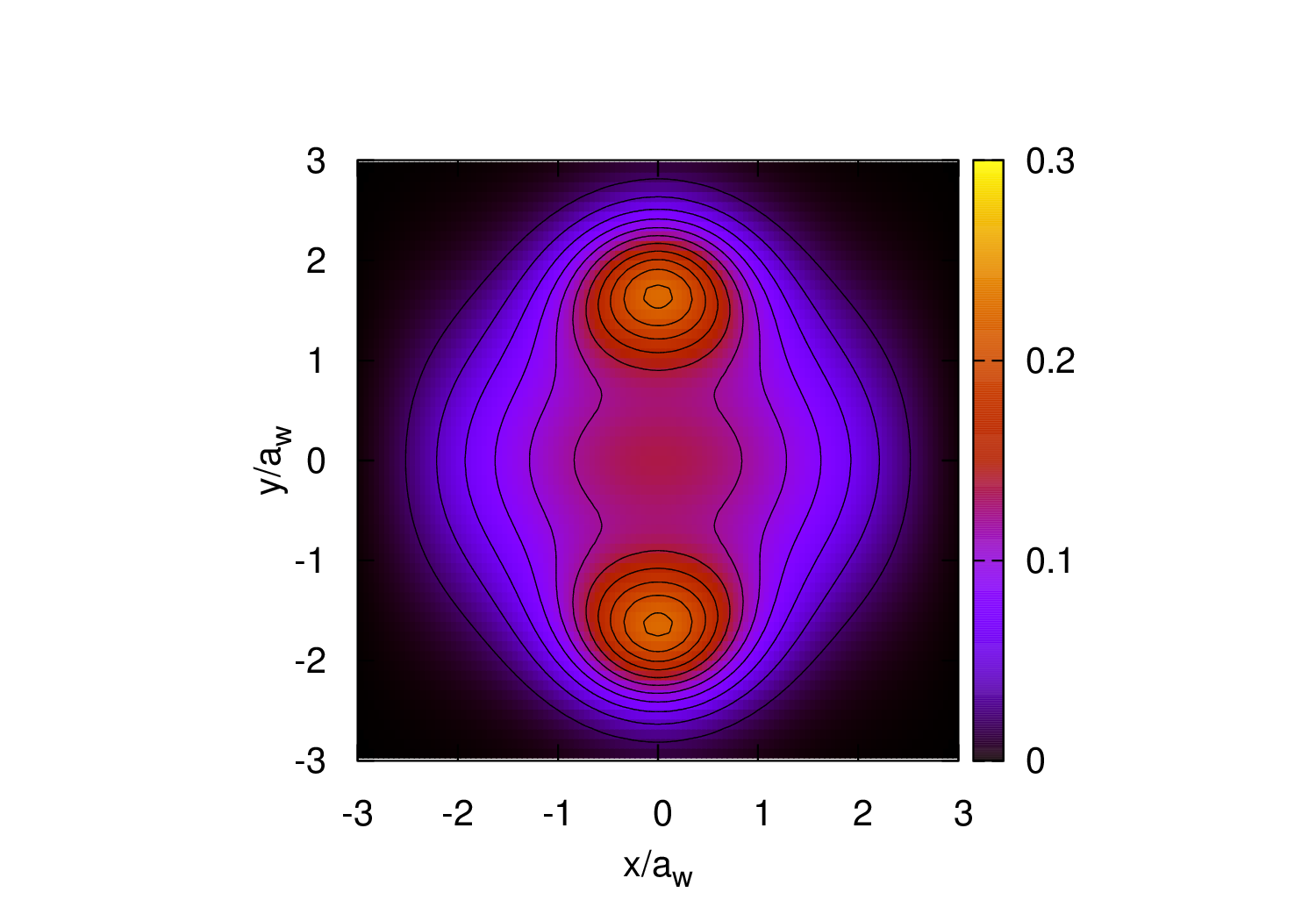}
      \caption{The probability density for 2 electrons and 1 photon with $g_\mathrm{EM}=0.05$ meV
               in the excited state $|\breve{61})$. $y$-polarized photon field, $\hbar\Omega_0=2.0$ meV, 
               and $B=0.1$ T.}
      \label{State-61}
\end{figure}

Yet, another way to gauge the effective Coulomb interaction is to excite the system by a
weak classical electrical dipole field polarized in either direction
\begin{equation}
      V_\mathrm{ext}(\mathbf{r},t)=V_t\, (\mathbf{r}\cdot\hat{\mathbf{e}})\exp{(-\Gamma t)}\sin{(\omega_1t)},
\end{equation}
with $V_t=0.1$ meV, $\hbar\Gamma = 0.1$ meV, and $\hbar\omega_1=0.4$ meV,
and $\hat{\mathbf{e}}$ determines the polarization of the pulse.
We are interested in the time evolution of the system after it is subjected to a
short dipole excitation pulse that can be polarized in the $x$- or $y$-direction and 
adds only a tiny amount of energy to it.\cite{arXiv:1502.06242} 
Since our system has a nontrivial geometry and interactions between the
electrons compared to the circular dot explored in an earlier publication with only Coulomb
interaction \cite{ANDP:ANDP201400048} we know that the system will respond not only with dipole
oscillations. There are higher order modes and Fig.\ \ref{Absorption}  
shows the Fourier transform of the expectation values for $x^2$ and $y^2$ for $y$-polarization
of the photon and the classical excitation field.
In both cases the fundamental modes (depicted in Fig.\ \ref{Absorption}) show a weak softening with
increasing coupling $g_\mathrm{EM}$. Since we are using a CI-formalism for the interaction we
can identify which many-body states participate in the oscillations, their occupation being shown
in the left panel of Fig.\ \ref{E-spectrum}. 
For low $g_\mathrm{EM}$ it is mainly $|\breve{61})$,
a state with peaks in the probability density located in the dots, but pushed a bit to higher values of
$y$ than in $|\breve{16})$ and then a considerably more density in the center between the dots, see 
Fig.\ \ref{State-61}.
Comparing the probability densities in the left panel of Fig.\ \ref{d0} and \ref{State-61} we
see that the weak dipole excitation of the system polarized in the $y$-direction causes 
internal motion of the charge or density with the charge in each dot oscillating in antiphase.

Furthermore, and again due to the CI-formalism, when we compare the dispersion for the energy of
the states $|\breve{16})$  and $|\breve{61})$ in the right panel of Fig.\ \ref{E-spectrum} we
see that the energy difference of the two states decreases with increasing $g_\mathrm{EM}$
reflecting the softening of the modes in Fig.\ \ref{Absorption}.
From this analysis of the closed system excited by a weak classical dipole electric field
we conclude that the effective Coulomb
interaction is reduced by the photon field, contrary to our conclusion earlier
when no energy was added to the system.

%%%%%%%%%%%%%%%%%%%%%%%%%%%%%%%%%%%%%%%%%%%%%%
\section{Open system}
We now turn our attention to the open system.
To open up the system to semi-infinite left and right leads with parabolic confinement in the 
$y$-direction we introduce a time-dependent coupling $H_T(t)$ between the central double-dot 
system and the leads.\cite{Gudmundsson:2013.305,Gudmundsson12:1109.4728}
\begin{equation}
      H_\mathrm{T}(t)=\sum_{i,l}\chi (t)\int dq\;
      \left\{T^l_{qi}c_{ql}^\dagger d_i + (T^l_{qi})^*d_i^\dagger c_{ql} \right\},
\label{H_T}
\end{equation}
where $d_i$ annihilates an electron in the single-electron state of the 
the central system labelled with $i$, and $c_{ql}^\dagger$ creates
one in lead $l$ with momentum and subband index labelled by the composite index $q$.\cite{Gudmundsson:2013.305} 
The coupling $T^l_{qi}$ depends on the shape of the wavefunctions for each single-electron state
in the leads and the central system in the contact areas defined between the subsystems.
The time-dependence of the coupling is determined by the switching function\cite{Gudmundsson:2013.305}
\begin{equation}
      \chi (t)=\left(1-\frac{2}{e^{\alpha t}+1}\right)
\label{chi}
\end{equation}
for both leads with $\alpha = 0.3$ ps$^{-1}$.
The state dependence of the coupling tensor in the single-electron picture is carried on to 
the corresponding coupling tensors in the many-body Fock spaces used in the 
calculation.\cite{Gudmundsson:2013.305,Gudmundsson12:1109.4728}
The Liouville-von Neuman equation for the density operator of the total system is then
projected on the central system by tracing out variables pertaining 
to the leads.\cite{Nakajima58:948,Zwanzig60:1338} In this proceedure 
we extract an overall coupling constant $g_\mathrm{LR}a_w^{3/2}=0.5$ meV.
In the generalized master equation (GME) for the reduced density operator
\begin{equation}
      \dot{\rho_\mathrm{S}}(t) = -\frac{i}{\hbar}[H_\mathrm{S},\rho_\mathrm{S}(t)]
      -\frac{1}{\hbar}\int_0^t dt' K[t,t-t';\rho_\mathrm{S}(t')]
\end{equation}
we keep up to second order terms of the lead-system coupling $H_T$ in the integral kernel of the 
dissipation term 
\begin{align}
      K[t,s;\rho_\mathrm{S}(t')]=\mathrm{Tr}_\mathrm{LR}&
      \left\{\left[ H_\mathrm{T}(t),
      \left[ U(s)H_\mathrm{T}(t')U^+(s), 
      \right. \right. \right. \nonumber\\
      &\left. \left. \left. 
      U_\mathrm{S}(s)\rho_\mathrm{S}(t')U_\mathrm{S}^+(s)\rho_\mathrm{L}\rho_\mathrm{R}\right]
      \right]\right\},
\end{align}
where $U_\mathrm{S}(t)=\exp{\{-iH_\mathrm{S}t/\hbar\}}$ is the unitary time evolution operator for 
the closed system, and $U(t)$ is the corresponding operator for the uncoupled system and the leads.
$\mathrm{Tr}_\mathrm{LR}$ is the trace operator for variables of the left and right leads, and 
$\rho_\mathrm{L}$ and $\rho_\mathrm{L}$ are the density matrices for the leads before coupling
to the central system.
The time-dependent reduced density operator can be used to calculate mean values of measurables 
of the central system \cite{Gudmundsson:2013.305,Gudmundsson12:1109.4728} under the influence
of the leads, that serve as electron reservoirs. 
Often, only properties of the steady state are of interest to evaluate the conductance of the 
system, life-time of certain states, their shift or broadening. Here, we are interested in analyzing 
the interaction of electrons that is best explored in the transient regime where the non-Markovian
character of the GME is still significant.    

In the numerical calculations for the closed system we commonly need the size of
various bases to be on the order of few thousand states. Due to the complex structure
of the dissipation terms in the GME we have to limit the size of the final many-body
basis used for the transport calculation to 120-192 states. The time-evolution of the 
closed system up to $t=1200$ ps for one set of parameters needs 1-2 days on 16 CPU 
and 512 GPU threads, and for the open system up to $300$ ps it needs 5-7 days 
on 12-16 CPU threads. The parallelization of the 
dissipation terms of the GME is a difficult task and could definitely be improved.
All programs are written in FORTRAN 2008 with OpenMP to maximize their speed.    

\begin{figure}[htbq]
      \includegraphics[width=0.48\textwidth,angle=0,viewport=50 10 2300 1680,clip]{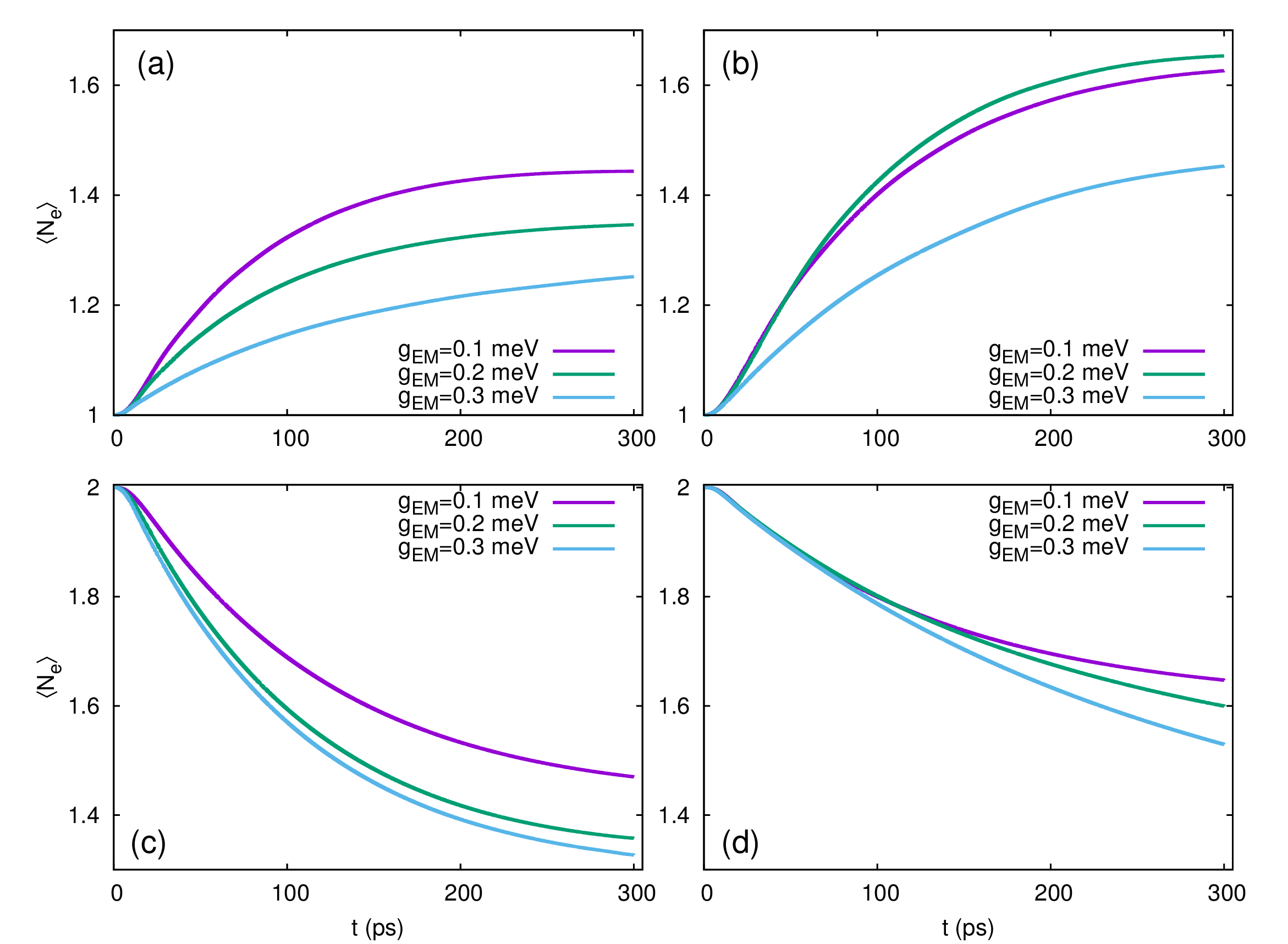}
      \caption{The average electron number $\langle N_e(t)\rangle$ as a function of time
               in an open system in the case of one electron initially in the first photon
               replica of the one-electron ground state in the system
               for $x$-polarized photon field (a), $y$-polarized photon field (b), and 
               for two electrons initially in the first photon replica of the two-electron
               ground state in the system for the case of $x$-polarized photon 
               field (c), $y$-polarized photon field (d). $\mu_L=1.4$ meV, $\mu_R=1.1$ meV,
               $g_{LR}a_w^{3/2}=0.5$ meV, $\hbar\Omega_0=2.0$ meV, $E_\mathrm{EM}=\hbar\omega=0.8$ meV, 
               and $B=0.1$ T.}
      \label{OpenNs}
\end{figure}

We begin the exploration of the open system by selecting the bias window to include the 
first photon replica ($|\breve{16})$) of the two-electron ground state ($|\breve{8})$).
The photon energy is $\hbar\omega =0.8$ meV, not close to a resonance for any low lying
2-electron states. Inspection of Fig.\ \ref{E-rof} shows that this photon
energy is not close to a resonance for any one-electron state for the case of $x$-polarization,
but close to a resonance for the one-electron ground state $|\breve{1})$ bound in the dots
and the one-electron state $|\breve{3})$ just above the dots in the case of $y$-polarization.
This can be seen by the fact that the mean photon number for $|\breve{3})$ and $|\breve{5})$
and their opposite spin counterparts is not close to an integer, reflecting a 
Rabi-splitting and entanglement.\cite{arXiv:1502.06242} The resulting extra polarizability of the 
one-electron states for the $y$-polarized photon field is not seen for the 
low energy 2-electron states. 
Figure \ref{OpenNs} shows the mean electron number in the system as a function of time.
In Fig.\ \ref{OpenNs}(c) and (d) we observe how one of the electrons leaves the system 
when initially the first photon replica of the two-electron ground state is occupied.
In panels (a) and (b) we see how a second electron enters the system when initially
the first photon replica ($|\breve{6})$) of the one-electron ground state ($|\breve{1})$) is occupied.   
The results in Fig.\ \ref{OpenNs} show that for the $x$-polarized photon field the depopulation
of the 2-electron states is enhanced with increasing electron-photon coupling $g_\mathrm{EM}$, and 
the addition of a second electron is resisted by increased $g_\mathrm{EM}$. The trend is not 
linear with $g_\mathrm{EM}$. Similar trend is seen for the $y$-polarization, but there the nonlinearity
is increased, and for some regime of low coupling it appears to be easier to add the second electron 
into the system. 

The conclusion is that in this situation (marked by the location of the bias with respect to the
energy spectrum and the geometry of the system)
the repulsive Coulomb interaction is enhanced by the dressing cavity photon field. 
An analysis of the polarization of the charge densities by the photon field gives supporting
arguments for this behavior. The dressing by the cavity-photon field changes the shape of 
the wavefunctions and the charge density in the contact areas to the 
leads,\cite{Gudmundsson09:113007} but due to the 
intricate interplay of the dressing by the para- and diamagnetic electron-photon terms
with the system geometry \cite{PhysRevB.49.8774,2040-8986-17-1-015201,arXiv:1502.06242}
the polarization of the charge densities is nontrivial, and in addition one has to account
for the changing occupation of the one- and two-electron states.

We have verified that qualitatively similar results are obtained for the two photon 
replica of the 2-electron ground state. The same can be said about the 2-electron 
ground state except there $\langle x^2\rangle$ decreases slightly with increasing   
$g_\mathrm{EM}$, and $\langle y^2\rangle$ only slightly increases. For the 2-electron
ground state the dispersion of the Fourier transform of $\langle x^2\rangle$ and 
$\langle y^2\rangle$ is fairly constant (corresponding to Fig.\ \ref{Absorption}), 
while the transport (corresponding to Fig.\ \ref{OpenNs}) indicates enhanced repulsion.

\section{Summary and discussion}
Summarizing, we emphasize that this analysis is only possible by taking account of
both the para- and diamagnetic electron-photon interactions and the geometry of the 
system through the wavefunctions used to construct the various bases used in the 
calculation.\cite{Gudmundsson:2013.305,PhysRevE.86.046701}
The parallel double dot system brings a
fine-tuned interplay between the para- and diamagnetic interactions that is not 
seen to the same extent in systems with a single dot,\cite{0953-8984-25-46-465302} 
or serial dots,\cite{Abdullah2014254} even though both interactions are important to
describe phenomena when the photon field is not in resonance with the electron 
system.\cite{PhysRevE.86.046701} To make this absolutely clear: Our results are
model dependent and point out a particular system with interesting properties. 

We notice that the answer to the question about the influence of the 
cavity photon mode on the effective electron-electron interaction depends on the stand point
we take to gauge it, or define it. The entity, two electrons dressed by cavity photons,
appears as a quasiparticle with different properties in different ``setups''.    

In our parallel double dot system investigated here
the dispersion of the energy of two-electron states bound to the dots shows that
the repulsion is enhanced by the electron-photon interaction. This is not a universal
behavior as can be seen for two electrons in a quantum wire without the embedded 
dots.\cite{PhysRevE.86.046701} We conclude thus that the effects of the photon
field in the cavity could not generally be described by a constant renormalization of
the dielectric constant of the surrounding material. 

We find that in our transport setup the photon field enhances the 
Coulomb repulsion of the two electrons in a parallel double quantum dot. This is true
for initial states below, in, or above the bias window, as the coupling to the 
leads at $t=0$ and the photon field enables slight coupling to states higher in the 
energy spectrum, and particularly to a one-electron state lower in the spectrum,
i.e.\ the one-photon replica of the one-electron ground state. 

In the closed system, in equilibrium, when gauged through the extent of the charge 
densities of the two-electron states, the effective interaction of the two electrons
is increased, or left untouched. 
On the other hand, when probed by observing the dispersion of the internal collective 
oscillation modes, excited by a weak classical electric dipole pulse, the effective 
interaction of the two electrons is reduced, or left unaffected. 
 
When we observe the strengthening of the repulsion between the 
two electrons due to the dressing of their mutual interaction
by the cavity photon field it is seen by the 
spreading of the charge (or polarization), but in the case of the 
transient transport we have to bear in mind the influence of occupation of states 
with one more or less electron than the initial state had.  

A clear exception to the enhancement of the repulsive interaction 
by the electron photon coupling in the present model is the softening of 
the relative component of the collective oscillation mode as measured by the Fourier transform
of the expectation values of $x^2$ and $y^2$ in Fig.\ \ref{Absorption}.
Here, Fig.\ \ref{State-61} compared to Fig.\ \ref{d0} indeed shows that
in addition to a spreading of the charge density both in $x$- and $y$-direction 
we see a very small accumulation of charge between the quantum dots during the oscillations.
This is valid for both linear polarization of the photon field in the 
$x$- and $y$-direction. The charge densities show a quadrupole oscillation
that is facilitated by the two terms of the electron-photon coupling thus
resulting in a softening of the mode with higher coupling.
 
The Fourier transform of the dipole response as measured by $\langle x\rangle$
and $\langle y\rangle$ is much less affected by the coupling giving weight to
our argument that the softening of the relative mode can be referred to the 
interaction between the electrons.  

One can wonder if the changes in the effective electron-electron interaction
observed here can be attributed to the Pauli principle and would thus also
be seen in noninteracting systems. We are sure the electron-photon interaction
would have effects on electrons that were not Coulomb interacting, but that
could be explained as the emergence of an electron-electron interaction. 
Generally, it is impossible to distinguish effects caused by changes in the 
structure of the state space and the underlying wavefunctions. One might
even ask if such an attempt would depend on the basis used. 
An increasing complexity of a state space can be used to describe the geometry 
of a system.

The system we explore here is a complex system with many parameters requiring
long CPU-time on parallel facilities. We find dispersion of the energy of the 
two-electron states indicating increased repulsion with stronger electron-photon 
coupling. As mentioned before this is not always the case for two-electron
systems in a photon cavity.\cite{PhysRevE.86.046701} We are not only gauging 
these systems in their equilibrium and it is likely that variants with different 
geometry can sustain different collective modes leading to a reduction of the effective 
electron-electron interaction as we have found here. 

It is interesting to notice that at two different points in the model we
are witnessing an effective light-light scattering or interactions.
First, through the dressing of the electrons by the cavity photons as
the Coulomb interaction of the electrons is also mediated by photons.
Second, the diamagnetic electron-photon interaction necessary here 
is essentially a nonrelativistic light-light interaction that in solid
state systems can lead to nonlinear effects for much lower strength than
is needed in a vacuum treated relativistically.   

The steady state in the open system only depends on the location of the bias
window in the energy spectrum of the system and the coupling strength.
It would thus be independent of the initial number of photons in the cavity.
Our results for the transient regime depend on the initial number of photons and 
electrons in addition to the location of the bias window, though we have verified
that our finding here is valid for an initial state with few photons (0-2).
We are describing a system with electron states, coupled to photons,
that can have very different coupling to the leads due to their geometry making the 
estimation of the time needed to reach a steady state difficult. This together with
the fact that conceivably some quantum electro-optical devices might be operated in 
the late transient time in order to save time and reduce dephasing lets us focus on 
that regime.

\begin{acknowledgments}
We acknowledge discussion with  C\u{a}t\u{a}lin Pa\c{s}cu Moca. 
This work was financially supported by the Research Fund of the University of Iceland,
and the Icelandic Instruments Fund. We also acknowledge support from the computational 
facilities of the Nordic High Performance Computing (NHPC), the Nordic network
NANOCONTROL, project No.: P-13053, and the Ministry of Science and Technology, Taiwan 
through Contract No. MOST 103-2112-M-239-001-MY3.
\end{acknowledgments}

%
%---------------------------------------------
%
%
\bibliographystyle{apsrev4-1}

\begin{thebibliography}{33}%
\makeatletter
\providecommand \@ifxundefined [1]{%
 \@ifx{#1\undefined}
}%
\providecommand \@ifnum [1]{%
 \ifnum #1\expandafter \@firstoftwo
 \else \expandafter \@secondoftwo
 \fi
}%
\providecommand \@ifx [1]{%
 \ifx #1\expandafter \@firstoftwo
 \else \expandafter \@secondoftwo
 \fi
}%
\providecommand \natexlab [1]{#1}%
\providecommand \enquote  [1]{``#1''}%
\providecommand \bibnamefont  [1]{#1}%
\providecommand \bibfnamefont [1]{#1}%
\providecommand \citenamefont [1]{#1}%
\providecommand \href@noop [0]{\@secondoftwo}%
\providecommand \href [0]{\begingroup \@sanitize@url \@href}%
\providecommand \@href[1]{\@@startlink{#1}\@@href}%
\providecommand \@@href[1]{\endgroup#1\@@endlink}%
\providecommand \@sanitize@url [0]{\catcode `\\12\catcode `\$12\catcode
  `\&12\catcode `\#12\catcode `\^12\catcode `\_12\catcode `\%12\relax}%
\providecommand \@@startlink[1]{}%
\providecommand \@@endlink[0]{}%
\providecommand \url  [0]{\begingroup\@sanitize@url \@url }%
\providecommand \@url [1]{\endgroup\@href {#1}{\urlprefix }}%
\providecommand \urlprefix  [0]{URL }%
\providecommand \Eprint [0]{\href }%
\providecommand \doibase [0]{http://dx.doi.org/}%
\providecommand \selectlanguage [0]{\@gobble}%
\providecommand \bibinfo  [0]{\@secondoftwo}%
\providecommand \bibfield  [0]{\@secondoftwo}%
\providecommand \translation [1]{[#1]}%
\providecommand \BibitemOpen [0]{}%
\providecommand \bibitemStop [0]{}%
\providecommand \bibitemNoStop [0]{.\EOS\space}%
\providecommand \EOS [0]{\spacefactor3000\relax}%
\providecommand \BibitemShut  [1]{\csname bibitem#1\endcsname}%
\let\auto@bib@innerbib\@empty
%</preamble>
\bibitem [{\citenamefont {Goy}\ \emph {et~al.}(1983)\citenamefont {Goy},
  \citenamefont {Raimond}, \citenamefont {Gross},\ and\ \citenamefont
  {Haroche}}]{PhysRevLett.50.1903}%
  \BibitemOpen
  \bibfield  {author} {\bibinfo {author} {\bibfnamefont {P.}~\bibnamefont
  {Goy}}, \bibinfo {author} {\bibfnamefont {J.~M.}\ \bibnamefont {Raimond}},
  \bibinfo {author} {\bibfnamefont {M.}~\bibnamefont {Gross}}, \ and\ \bibinfo
  {author} {\bibfnamefont {S.}~\bibnamefont {Haroche}},\ }\href {\doibase
  10.1103/PhysRevLett.50.1903} {\bibfield  {journal} {\bibinfo  {journal}
  {Phys. Rev. Lett.}\ }\textbf {\bibinfo {volume} {50}},\ \bibinfo {pages}
  {1903} (\bibinfo {year} {1983})}\BibitemShut {NoStop}%
\bibitem [{\citenamefont {Petersson}\ \emph {et~al.}(2012)\citenamefont
  {Petersson}, \citenamefont {McFaul}, \citenamefont {Schroer}, \citenamefont
  {Jung}, \citenamefont {Taylor}, \citenamefont {Houck},\ and\ \citenamefont
  {Petta}}]{Peterson380:2012}%
  \BibitemOpen
  \bibfield  {author} {\bibinfo {author} {\bibfnamefont {K.~D.}\ \bibnamefont
  {Petersson}}, \bibinfo {author} {\bibfnamefont {L.~W.}\ \bibnamefont
  {McFaul}}, \bibinfo {author} {\bibfnamefont {M.~D.}\ \bibnamefont {Schroer}},
  \bibinfo {author} {\bibfnamefont {M.}~\bibnamefont {Jung}}, \bibinfo {author}
  {\bibfnamefont {J.~M.}\ \bibnamefont {Taylor}}, \bibinfo {author}
  {\bibfnamefont {A.~A.}\ \bibnamefont {Houck}}, \ and\ \bibinfo {author}
  {\bibfnamefont {J.~R.}\ \bibnamefont {Petta}},\ }\href@noop {} {\bibfield
  {journal} {\bibinfo  {journal} {Nature}\ ,\ \bibinfo {pages} {380}} (\bibinfo
  {year} {2012})}\BibitemShut {NoStop}%
\bibitem [{\citenamefont {Maragkou}\ \emph {et~al.}(2012)\citenamefont
  {Maragkou}, \citenamefont {Nowak}, \citenamefont {Gallardo}, \citenamefont
  {van~der Meulen}, \citenamefont {Prieto}, \citenamefont {Martinez},
  \citenamefont {Postigo},\ and\ \citenamefont {Calleja}}]{PhysRevB.86.085316}%
  \BibitemOpen
  \bibfield  {author} {\bibinfo {author} {\bibfnamefont {M.}~\bibnamefont
  {Maragkou}}, \bibinfo {author} {\bibfnamefont {A.~K.}\ \bibnamefont {Nowak}},
  \bibinfo {author} {\bibfnamefont {E.}~\bibnamefont {Gallardo}}, \bibinfo
  {author} {\bibfnamefont {H.~P.}\ \bibnamefont {van~der Meulen}}, \bibinfo
  {author} {\bibfnamefont {I.}~\bibnamefont {Prieto}}, \bibinfo {author}
  {\bibfnamefont {L.~J.}\ \bibnamefont {Martinez}}, \bibinfo {author}
  {\bibfnamefont {P.~A.}\ \bibnamefont {Postigo}}, \ and\ \bibinfo {author}
  {\bibfnamefont {J.~M.}\ \bibnamefont {Calleja}},\ }\href {\doibase
  10.1103/PhysRevB.86.085316} {\bibfield  {journal} {\bibinfo  {journal} {Phys.
  Rev. B}\ }\textbf {\bibinfo {volume} {86}},\ \bibinfo {pages} {085316}
  (\bibinfo {year} {2012})}\BibitemShut {NoStop}%
\bibitem [{\citenamefont {Gallardo}\ \emph {et~al.}(2010)\citenamefont
  {Gallardo}, \citenamefont {Martínez}, \citenamefont {Nowak}, \citenamefont
  {van~der Meulen}, \citenamefont {Calleja}, \citenamefont {Tejedor},
  \citenamefont {Prieto}, \citenamefont {Granados}, \citenamefont {Taboada},
  \citenamefont {García},\ and\ \citenamefont {Postigo}}]{Gallardo:10}%
  \BibitemOpen
  \bibfield  {author} {\bibinfo {author} {\bibfnamefont {E.}~\bibnamefont
  {Gallardo}}, \bibinfo {author} {\bibfnamefont {L.~J.}\ \bibnamefont
  {Martínez}}, \bibinfo {author} {\bibfnamefont {A.~K.}\ \bibnamefont
  {Nowak}}, \bibinfo {author} {\bibfnamefont {H.~P.}\ \bibnamefont {van~der
  Meulen}}, \bibinfo {author} {\bibfnamefont {J.~M.}\ \bibnamefont {Calleja}},
  \bibinfo {author} {\bibfnamefont {C.}~\bibnamefont {Tejedor}}, \bibinfo
  {author} {\bibfnamefont {I.}~\bibnamefont {Prieto}}, \bibinfo {author}
  {\bibfnamefont {D.}~\bibnamefont {Granados}}, \bibinfo {author}
  {\bibfnamefont {A.~G.}\ \bibnamefont {Taboada}}, \bibinfo {author}
  {\bibfnamefont {J.~M.}\ \bibnamefont {García}}, \ and\ \bibinfo {author}
  {\bibfnamefont {P.~A.}\ \bibnamefont {Postigo}},\ }\href {\doibase
  10.1364/OE.18.013301} {\bibfield  {journal} {\bibinfo  {journal} {Opt.
  Express}\ }\textbf {\bibinfo {volume} {18}},\ \bibinfo {pages} {13301}
  (\bibinfo {year} {2010})}\BibitemShut {NoStop}%
\bibitem [{\citenamefont {Frey}\ \emph {et~al.}(2012)\citenamefont {Frey},
  \citenamefont {Leek}, \citenamefont {Beck}, \citenamefont {Blais},
  \citenamefont {Ihn}, \citenamefont {Ensslin},\ and\ \citenamefont
  {Wallraff}}]{Frey11:01}%
  \BibitemOpen
  \bibfield  {author} {\bibinfo {author} {\bibfnamefont {T.}~\bibnamefont
  {Frey}}, \bibinfo {author} {\bibfnamefont {P.~J.}\ \bibnamefont {Leek}},
  \bibinfo {author} {\bibfnamefont {M.}~\bibnamefont {Beck}}, \bibinfo {author}
  {\bibfnamefont {A.}~\bibnamefont {Blais}}, \bibinfo {author} {\bibfnamefont
  {T.}~\bibnamefont {Ihn}}, \bibinfo {author} {\bibfnamefont {K.}~\bibnamefont
  {Ensslin}}, \ and\ \bibinfo {author} {\bibfnamefont {A.}~\bibnamefont
  {Wallraff}},\ }\href {\doibase 10.1103/PhysRevLett.108.046807} {\bibfield
  {journal} {\bibinfo  {journal} {Phys. Rev. Lett.}\ }\textbf {\bibinfo
  {volume} {108}},\ \bibinfo {pages} {046807} (\bibinfo {year}
  {2012})}\BibitemShut {NoStop}%
\bibitem [{\citenamefont {Sturm}\ \emph {et~al.}(2011)\citenamefont {Sturm},
  \citenamefont {Hilmer}, \citenamefont {Rheinl{\"a}nder}, \citenamefont
  {Schmidt-Grund},\ and\ \citenamefont {Grundmann}}]{PhysRevB.83.205301}%
  \BibitemOpen
  \bibfield  {author} {\bibinfo {author} {\bibfnamefont {C.}~\bibnamefont
  {Sturm}}, \bibinfo {author} {\bibfnamefont {H.}~\bibnamefont {Hilmer}},
  \bibinfo {author} {\bibfnamefont {B.}~\bibnamefont {Rheinl{\"a}nder}}, \bibinfo
  {author} {\bibfnamefont {R.}~\bibnamefont {Schmidt-Grund}}, \ and\ \bibinfo
  {author} {\bibfnamefont {M.}~\bibnamefont {Grundmann}},\ }\href {\doibase
  10.1103/PhysRevB.83.205301} {\bibfield  {journal} {\bibinfo  {journal} {Phys.
  Rev. B}\ }\textbf {\bibinfo {volume} {83}},\ \bibinfo {pages} {205301}
  (\bibinfo {year} {2011})}\BibitemShut {NoStop}%
\bibitem [{\citenamefont {Bergenfeldt}\ and\ \citenamefont
  {Samuelsson}(2013)}]{PhysRevB.87.195427}%
  \BibitemOpen
  \bibfield  {author} {\bibinfo {author} {\bibfnamefont {C.}~\bibnamefont
  {Bergenfeldt}}\ and\ \bibinfo {author} {\bibfnamefont {P.}~\bibnamefont
  {Samuelsson}},\ }\href {\doibase 10.1103/PhysRevB.87.195427} {\bibfield
  {journal} {\bibinfo  {journal} {Phys. Rev. B}\ }\textbf {\bibinfo {volume}
  {87}},\ \bibinfo {pages} {195427} (\bibinfo {year} {2013})}\BibitemShut
  {NoStop}%
\bibitem [{\citenamefont {Ates}\ \emph {et~al.}(2009)\citenamefont {Ates},
  \citenamefont {Ulrich}, \citenamefont {Ulhaq}, \citenamefont {Reitzenstein},
  \citenamefont {L{\"o}ffler}, \citenamefont {H{\"o}fling}, \citenamefont {Forchel},\
  and\ \citenamefont {Michler}}]{ates2009:724}%
  \BibitemOpen
  \bibfield  {author} {\bibinfo {author} {\bibfnamefont {S.}~\bibnamefont
  {Ates}}, \bibinfo {author} {\bibfnamefont {S.~M.}\ \bibnamefont {Ulrich}},
  \bibinfo {author} {\bibfnamefont {A.}~\bibnamefont {Ulhaq}}, \bibinfo
  {author} {\bibfnamefont {S.}~\bibnamefont {Reitzenstein}}, \bibinfo {author}
  {\bibfnamefont {A.}~\bibnamefont {L{\"o}ffler}}, \bibinfo {author}
  {\bibfnamefont {S.}~\bibnamefont {H{\"o}fling}}, \bibinfo {author}
  {\bibfnamefont {A.}~\bibnamefont {Forchel}}, \ and\ \bibinfo {author}
  {\bibnamefont {Michler}},\ }\href@noop {} {\bibfield  {journal} {\bibinfo
  {journal} {Nature Photonics}\ }\textbf {\bibinfo {volume} {3}},\ \bibinfo
  {pages} {724} (\bibinfo {year} {2009})}\BibitemShut {NoStop}%
\bibitem [{\citenamefont {Majumdar}\ \emph {et~al.}(2010)\citenamefont
  {Majumdar}, \citenamefont {Faraon}, \citenamefont {Kim}, \citenamefont
  {Englund}, \citenamefont {Kim}, \citenamefont {Petroff},\ and\ \citenamefont
  {{Vu\u{c}kovi\'{c}}}}]{PhysRevB.82.045306}%
  \BibitemOpen
  \bibfield  {author} {\bibinfo {author} {\bibfnamefont {A.}~\bibnamefont
  {Majumdar}}, \bibinfo {author} {\bibfnamefont {A.}~\bibnamefont {Faraon}},
  \bibinfo {author} {\bibfnamefont {E.~D.}\ \bibnamefont {Kim}}, \bibinfo
  {author} {\bibfnamefont {D.}~\bibnamefont {Englund}}, \bibinfo {author}
  {\bibfnamefont {H.}~\bibnamefont {Kim}}, \bibinfo {author} {\bibfnamefont
  {P.}~\bibnamefont {Petroff}}, \ and\ \bibinfo {author} {\bibfnamefont
  {J.}~\bibnamefont {{Vu\u{c}kovi\'{c}}}},\ }\href {\doibase 10.1103/PhysRevB.82.045306}
  {\bibfield  {journal} {\bibinfo  {journal} {Phys. Rev. B}\ }\textbf {\bibinfo
  {volume} {82}},\ \bibinfo {pages} {045306} (\bibinfo {year}
  {2010})}\BibitemShut {NoStop}%
\bibitem [{\citenamefont {Rammer}\ and\ \citenamefont
  {Smith}(1986)}]{RevModPhys.58.323}%
  \BibitemOpen
  \bibfield  {author} {\bibinfo {author} {\bibfnamefont {J.}~\bibnamefont
  {Rammer}}\ and\ \bibinfo {author} {\bibfnamefont {H.}~\bibnamefont {Smith}},\
  }\href {\doibase 10.1103/RevModPhys.58.323} {\bibfield  {journal} {\bibinfo
  {journal} {Rev. Mod. Phys.}\ }\textbf {\bibinfo {volume} {58}},\ \bibinfo
  {pages} {323–359} (\bibinfo {year} {1986})}\BibitemShut {NoStop}%
\bibitem [{\citenamefont {Datta}(2000)}]{Datta00:253}%
  \BibitemOpen
  \bibfield  {author} {\bibinfo {author} {\bibfnamefont {S.}~\bibnamefont
  {Datta}},\ }\href@noop {} {\bibfield  {journal} {\bibinfo  {journal}
  {Superlattices and Microstructures}\ }\textbf {\bibinfo {volume} {28}},\
  \bibinfo {pages} {253} (\bibinfo {year} {2000})}\BibitemShut {NoStop}%
\bibitem [{\citenamefont {Harbola}\ \emph {et~al.}(2006)\citenamefont
  {Harbola}, \citenamefont {Esposito},\ and\ \citenamefont
  {Mukamel}}]{Harbola06:235309}%
  \BibitemOpen
  \bibfield  {author} {\bibinfo {author} {\bibfnamefont {U.}~\bibnamefont
  {Harbola}}, \bibinfo {author} {\bibfnamefont {M.}~\bibnamefont {Esposito}}, \
  and\ \bibinfo {author} {\bibfnamefont {S.}~\bibnamefont {Mukamel}},\ }\href
  {http://link.aps.org/abstract/PRB/v74/e235309} {\bibfield  {journal}
  {\bibinfo  {journal} {Phys. Rev. B}\ }\textbf {\bibinfo {volume} {74}},\
  \bibinfo {pages} {235309} (\bibinfo {year} {2006})}\BibitemShut {NoStop}%
\bibitem [{\citenamefont {Timm}(2008)}]{Timm08:195416}%
  \BibitemOpen
  \bibfield  {author} {\bibinfo {author} {\bibfnamefont {C.}~\bibnamefont
  {Timm}},\ }\href@noop {} {\bibfield  {journal} {\bibinfo  {journal} {Phys.
  Rev. B}\ }\textbf {\bibinfo {volume} {77}},\ \bibinfo {pages} {195416}
  (\bibinfo {year} {2008})}\BibitemShut {NoStop}%
\bibitem [{\citenamefont {Moldoveanu}\ \emph {et~al.}(2010)\citenamefont
  {Moldoveanu}, \citenamefont {Manolescu}, \citenamefont {Tang},\ and\
  \citenamefont {Gudmundsson}}]{Moldoveanu10:155442}%
  \BibitemOpen
  \bibfield  {author} {\bibinfo {author} {\bibfnamefont {V.}~\bibnamefont
  {Moldoveanu}}, \bibinfo {author} {\bibfnamefont {A.}~\bibnamefont
  {Manolescu}}, \bibinfo {author} {\bibfnamefont {C.-S.}\ \bibnamefont {Tang}},
  \ and\ \bibinfo {author} {\bibfnamefont {V.}~\bibnamefont {Gudmundsson}},\
  }\href {http://link.aps.org/doi/10.1103/PhysRevB.81.155442} {\bibfield
  {journal} {\bibinfo  {journal} {Phys. Rev. B}\ }\textbf {\bibinfo {volume}
  {81}},\ \bibinfo {pages} {155442} (\bibinfo {year} {2010})}\BibitemShut
  {NoStop}%
\bibitem [{\citenamefont {Pfannkuche}\ and\ \citenamefont
  {Gerhardts}(1991)}]{Pfannkuche91:13132}%
  \BibitemOpen
  \bibfield  {author} {\bibinfo {author} {\bibfnamefont {D.}~\bibnamefont
  {Pfannkuche}}\ and\ \bibinfo {author} {\bibfnamefont {R.}~\bibnamefont
  {Gerhardts}},\ }\href@noop {} {\bibfield  {journal} {\bibinfo  {journal}
  {Phys. Rev. B}\ }\textbf {\bibinfo {volume} {44}},\ \bibinfo {pages} {13132}
  (\bibinfo {year} {1991})}\BibitemShut {NoStop}%
\bibitem [{\citenamefont {Maksym}\ and\ \citenamefont
  {Chakraborty}(1990)}]{Maksym90:108}%
  \BibitemOpen
  \bibfield  {author} {\bibinfo {author} {\bibfnamefont {P.~A.}\ \bibnamefont
  {Maksym}}\ and\ \bibinfo {author} {\bibfnamefont {T.}~\bibnamefont
  {Chakraborty}},\ }\href@noop {} {\bibfield  {journal} {\bibinfo  {journal}
  {Phys. Rev. Lett}\ }\textbf {\bibinfo {volume} {65}},\ \bibinfo {pages} {108}
  (\bibinfo {year} {1990})}\BibitemShut {NoStop}%
\bibitem [{\citenamefont {Kibis}\ \emph {et~al.}(2014)\citenamefont {Kibis},
  \citenamefont {Arnardottir},\ and\ \citenamefont
  {Shelykh}}]{PhysRevA.90.055802}%
  \BibitemOpen
  \bibfield  {author} {\bibinfo {author} {\bibfnamefont {O.~V.}\ \bibnamefont
  {Kibis}}, \bibinfo {author} {\bibfnamefont {K.~B.}\ \bibnamefont
  {Arnardottir}}, \ and\ \bibinfo {author} {\bibfnamefont {I.~A.}\ \bibnamefont
  {Shelykh}},\ }\href {\doibase 10.1103/PhysRevA.90.055802} {\bibfield
  {journal} {\bibinfo  {journal} {Phys. Rev. A}\ }\textbf {\bibinfo {volume}
  {90}},\ \bibinfo {pages} {055802} (\bibinfo {year} {2014})}\BibitemShut
  {NoStop}%
\bibitem [{\citenamefont {Gudmundsson}\ \emph {et~al.}(2013)\citenamefont
  {Gudmundsson}, \citenamefont {Jonasson}, \citenamefont {Arnold},
  \citenamefont {Tang}, \citenamefont {Goan},\ and\ \citenamefont
  {Manolescu}}]{Gudmundsson:2013.305}%
  \BibitemOpen
  \bibfield  {author} {\bibinfo {author} {\bibfnamefont {V.}~\bibnamefont
  {Gudmundsson}}, \bibinfo {author} {\bibfnamefont {O.}~\bibnamefont
  {Jonasson}}, \bibinfo {author} {\bibfnamefont {T.}~\bibnamefont {Arnold}},
  \bibinfo {author} {\bibfnamefont {C.-S.}\ \bibnamefont {Tang}}, \bibinfo
  {author} {\bibfnamefont {H.-S.}\ \bibnamefont {Goan}}, \ and\ \bibinfo
  {author} {\bibfnamefont {A.}~\bibnamefont {Manolescu}},\ }\href {\doibase
  10.1002/prop.201200053} {\bibfield  {journal} {\bibinfo  {journal}
  {Fortschritte der Physik}\ }\textbf {\bibinfo {volume} {61}},\ \bibinfo
  {pages} {305} (\bibinfo {year} {2013})}\BibitemShut {NoStop}%
\bibitem [{\citenamefont {Jaynes}\ and\ \citenamefont
  {Cummings}(1963)}]{Jaynes63:89}%
  \BibitemOpen
  \bibfield  {author} {\bibinfo {author} {\bibfnamefont {E.~T.}\ \bibnamefont
  {Jaynes}}\ and\ \bibinfo {author} {\bibfnamefont {F.~W.}\ \bibnamefont
  {Cummings}},\ }\href
  {http://bayes.wustl.edu/etj/articles/comparison.qm.scr.pdf} {\bibfield
  {journal} {\bibinfo  {journal} {Proc. IEEE.}\ }\textbf {\bibinfo {volume}
  {51}},\ \bibinfo {pages} {89} (\bibinfo {year} {1963})}\BibitemShut {NoStop}%
\bibitem [{\citenamefont {Feranchuk}\ \emph {et~al.}(1996)\citenamefont
  {Feranchuk}, \citenamefont {Komarov},\ and\ \citenamefont
  {Ulyanenkov}}]{Feranchuk96:4035}%
  \BibitemOpen
  \bibfield  {author} {\bibinfo {author} {\bibfnamefont {I.~D.}\ \bibnamefont
  {Feranchuk}}, \bibinfo {author} {\bibfnamefont {L.~I.}\ \bibnamefont
  {Komarov}}, \ and\ \bibinfo {author} {\bibfnamefont {A.~P.}\ \bibnamefont
  {Ulyanenkov}},\ }\href {http://iopscience.iop.org/0305-4470/29/14/026}
  {\bibfield  {journal} {\bibinfo  {journal} {J. Phys. A: Math. Gen.}\ }\textbf
  {\bibinfo {volume} {29}},\ \bibinfo {pages} {4035} (\bibinfo {year}
  {1996})}\BibitemShut {NoStop}%
\bibitem [{\citenamefont {Jonasson}\ \emph
  {et~al.}(2012{\natexlab{a}})\citenamefont {Jonasson}, \citenamefont {Tang},
  \citenamefont {Goan}, \citenamefont {Manolescu},\ and\ \citenamefont
  {Gudmundsson}}]{Jonasson2011:01}%
  \BibitemOpen
  \bibfield  {author} {\bibinfo {author} {\bibfnamefont {O.}~\bibnamefont
  {Jonasson}}, \bibinfo {author} {\bibfnamefont {C.-S.}\ \bibnamefont {Tang}},
  \bibinfo {author} {\bibfnamefont {H.-S.}\ \bibnamefont {Goan}}, \bibinfo
  {author} {\bibfnamefont {A.}~\bibnamefont {Manolescu}}, \ and\ \bibinfo
  {author} {\bibfnamefont {V.}~\bibnamefont {Gudmundsson}},\ }\href
  {http://stacks.iop.org/1367-2630/14/i=1/a=013036} {\bibfield  {journal}
  {\bibinfo  {journal} {New Journal of Physics}\ }\textbf {\bibinfo {volume}
  {14}},\ \bibinfo {pages} {013036} (\bibinfo {year}
  {2012}{\natexlab{a}})}\BibitemShut {NoStop}%
\bibitem [{\citenamefont {Jonasson}\ \emph
  {et~al.}(2012{\natexlab{b}})\citenamefont {Jonasson}, \citenamefont {Tang},
  \citenamefont {Goan}, \citenamefont {Manolescu},\ and\ \citenamefont
  {Gudmundsson}}]{PhysRevE.86.046701}%
  \BibitemOpen
  \bibfield  {author} {\bibinfo {author} {\bibfnamefont {O.}~\bibnamefont
  {Jonasson}}, \bibinfo {author} {\bibfnamefont {C.-S.}\ \bibnamefont {Tang}},
  \bibinfo {author} {\bibfnamefont {H.-S.}\ \bibnamefont {Goan}}, \bibinfo
  {author} {\bibfnamefont {A.}~\bibnamefont {Manolescu}}, \ and\ \bibinfo
  {author} {\bibfnamefont {V.}~\bibnamefont {Gudmundsson}},\ }\href {\doibase
  10.1103/PhysRevE.86.046701} {\bibfield  {journal} {\bibinfo  {journal} {Phys.
  Rev. E}\ }\textbf {\bibinfo {volume} {86}},\ \bibinfo {pages} {046701}
  (\bibinfo {year} {2012}{\natexlab{b}})}\BibitemShut {NoStop}%
\bibitem [{\citenamefont {Arnold}\ \emph {et~al.}(2015)\citenamefont {Arnold},
  \citenamefont {Tang}, \citenamefont {Manolescu},\ and\ \citenamefont
  {Gudmundsson}}]{2040-8986-17-1-015201}%
  \BibitemOpen
  \bibfield  {author} {\bibinfo {author} {\bibfnamefont {T.}~\bibnamefont
  {Arnold}}, \bibinfo {author} {\bibfnamefont {C.-S.}\ \bibnamefont {Tang}},
  \bibinfo {author} {\bibfnamefont {A.}~\bibnamefont {Manolescu}}, \ and\
  \bibinfo {author} {\bibfnamefont {V.}~\bibnamefont {Gudmundsson}},\ }\href
  {http://stacks.iop.org/2040-8986/17/i=1/a=015201} {\bibfield  {journal}
  {\bibinfo  {journal} {Journal of Optics}\ }\textbf {\bibinfo {volume} {17}},\
  \bibinfo {pages} {015201} (\bibinfo {year} {2015})}\BibitemShut {NoStop}%
\bibitem [{\citenamefont {Chen}\ \emph {et~al.}(2011)\citenamefont {Chen},
  \citenamefont {Liu}, \citenamefont {Zhang},\ and\ \citenamefont
  {Wang}}]{0295-5075-96-1-14003}%
  \BibitemOpen
  \bibfield  {author} {\bibinfo {author} {\bibfnamefont {Q.-H.}\ \bibnamefont
  {Chen}}, \bibinfo {author} {\bibfnamefont {T.}~\bibnamefont {Liu}}, \bibinfo
  {author} {\bibfnamefont {Y.-Y.}\ \bibnamefont {Zhang}}, \ and\ \bibinfo
  {author} {\bibfnamefont {K.-L.}\ \bibnamefont {Wang}},\ }\href
  {http://stacks.iop.org/0295-5075/96/i=1/a=14003} {\bibfield  {journal}
  {\bibinfo  {journal} {EPL (Europhysics Letters)}\ }\textbf {\bibinfo {volume}
  {96}},\ \bibinfo {pages} {14003} (\bibinfo {year} {2011})}\BibitemShut
  {NoStop}%
\bibitem [{\citenamefont {Gudmundsson}\ \emph {et~al.}(2012)\citenamefont
  {Gudmundsson}, \citenamefont {Jonasson}, \citenamefont {Tang}, \citenamefont
  {Goan},\ and\ \citenamefont {Manolescu}}]{Gudmundsson12:1109.4728}%
  \BibitemOpen
  \bibfield  {author} {\bibinfo {author} {\bibfnamefont {V.}~\bibnamefont
  {Gudmundsson}}, \bibinfo {author} {\bibfnamefont {O.}~\bibnamefont
  {Jonasson}}, \bibinfo {author} {\bibfnamefont {C.-S.}\ \bibnamefont {Tang}},
  \bibinfo {author} {\bibfnamefont {H.-S.}\ \bibnamefont {Goan}}, \ and\
  \bibinfo {author} {\bibfnamefont {A.}~\bibnamefont {Manolescu}},\ }\href
  {\doibase 10.1103/PhysRevB.85.075306} {\bibfield  {journal} {\bibinfo
  {journal} {Phys. Rev. B}\ }\textbf {\bibinfo {volume} {85}},\ \bibinfo
  {pages} {075306} (\bibinfo {year} {2012})}\BibitemShut {NoStop}%
\bibitem [{\citenamefont {Gudmundsson}\ \emph {et~al.}(2015)\citenamefont
  {Gudmundsson}, \citenamefont {Sitek}, \citenamefont {yi~Lin}, \citenamefont
  {Abdullah}, \citenamefont {Tang},\ and\ \citenamefont
  {Manolescu}}]{arXiv:1502.06242}%
  \BibitemOpen
  \bibfield  {author} {\bibinfo {author} {\bibfnamefont {V.}~\bibnamefont
  {Gudmundsson}}, \bibinfo {author} {\bibfnamefont {A.}~\bibnamefont {Sitek}},
  \bibinfo {author} {\bibfnamefont {P.}~\bibnamefont {yi~Lin}}, \bibinfo
  {author} {\bibfnamefont {N.~R.}\ \bibnamefont {Abdullah}}, \bibinfo {author}
  {\bibfnamefont {C.-S.}\ \bibnamefont {Tang}}, \ and\ \bibinfo {author}
  {\bibfnamefont {A.}~\bibnamefont {Manolescu}},\ }\href {\doibase
  10.1021/acsphotonics.5b00115} {\bibfield  {journal} {\bibinfo  {journal} {ACS
  Photonics}\ }\textbf {\bibinfo {volume} {2}},\ \bibinfo {pages} {930}
  (\bibinfo {year} {2015})}\BibitemShut {NoStop}%
\bibitem [{\citenamefont {Gudmundsson}\ \emph {et~al.}(2014)\citenamefont
  {Gudmundsson}, \citenamefont {Hauksson}, \citenamefont {Johnsen},
  \citenamefont {Reinisch}, \citenamefont {Manolescu}, \citenamefont {Besse},\
  and\ \citenamefont {Dujardin}}]{ANDP:ANDP201400048}%
  \BibitemOpen
  \bibfield  {author} {\bibinfo {author} {\bibfnamefont {V.}~\bibnamefont
  {Gudmundsson}}, \bibinfo {author} {\bibfnamefont {S.}~\bibnamefont
  {Hauksson}}, \bibinfo {author} {\bibfnamefont {A.}~\bibnamefont {Johnsen}},
  \bibinfo {author} {\bibfnamefont {G.}~\bibnamefont {Reinisch}}, \bibinfo
  {author} {\bibfnamefont {A.}~\bibnamefont {Manolescu}}, \bibinfo {author}
  {\bibfnamefont {C.}~\bibnamefont {Besse}}, \ and\ \bibinfo {author}
  {\bibfnamefont {G.}~\bibnamefont {Dujardin}},\ }\href {\doibase
  10.1002/andp.201400048} {\bibfield  {journal} {\bibinfo  {journal} {Annalen
  der Physik}\ }\textbf {\bibinfo {volume} {526}},\ \bibinfo {pages} {235}
  (\bibinfo {year} {2014})}\BibitemShut {NoStop}%
\bibitem [{\citenamefont {Nakajima}(1958)}]{Nakajima58:948}%
  \BibitemOpen
  \bibfield  {author} {\bibinfo {author} {\bibfnamefont {S.}~\bibnamefont
  {Nakajima}},\ }\href@noop {} {\bibfield  {journal} {\bibinfo  {journal}
  {Prog. Theor. Phys.}\ }\textbf {\bibinfo {volume} {20}},\ \bibinfo {pages}
  {948} (\bibinfo {year} {1958})}\BibitemShut {NoStop}%
\bibitem [{\citenamefont {Zwanzig}(1960)}]{Zwanzig60:1338}%
  \BibitemOpen
  \bibfield  {author} {\bibinfo {author} {\bibfnamefont {R.}~\bibnamefont
  {Zwanzig}},\ }\href@noop {} {\bibfield  {journal} {\bibinfo  {journal} {J.
  Chem. Phys.}\ }\textbf {\bibinfo {volume} {33}},\ \bibinfo {pages} {1338}
  (\bibinfo {year} {1960})}\BibitemShut {NoStop}%
\bibitem [{\citenamefont {Gudmundsson}\ \emph {et~al.}(2009)\citenamefont
  {Gudmundsson}, \citenamefont {Gainar}, \citenamefont {Tang}, \citenamefont
  {Moldoveanu},\ and\ \citenamefont {Manolescu}}]{Gudmundsson09:113007}%
  \BibitemOpen
  \bibfield  {author} {\bibinfo {author} {\bibfnamefont {V.}~\bibnamefont
  {Gudmundsson}}, \bibinfo {author} {\bibfnamefont {C.}~\bibnamefont {Gainar}},
  \bibinfo {author} {\bibfnamefont {C.-S.}\ \bibnamefont {Tang}}, \bibinfo
  {author} {\bibfnamefont {V.}~\bibnamefont {Moldoveanu}}, \ and\ \bibinfo
  {author} {\bibfnamefont {A.}~\bibnamefont {Manolescu}},\ }\href
  {http://stacks.iop.org/1367-2630/11/113007} {\bibfield  {journal} {\bibinfo
  {journal} {New Journal of Physics}\ }\textbf {\bibinfo {volume} {11}},\
  \bibinfo {pages} {113007} (\bibinfo {year} {2009})}\BibitemShut {NoStop}%
\bibitem [{\citenamefont {Savona}\ \emph {et~al.}(1994)\citenamefont {Savona},
  \citenamefont {Hradil}, \citenamefont {Quattropani},\ and\ \citenamefont
  {Schwendimann}}]{PhysRevB.49.8774}%
  \BibitemOpen
  \bibfield  {author} {\bibinfo {author} {\bibfnamefont {V.}~\bibnamefont
  {Savona}}, \bibinfo {author} {\bibfnamefont {Z.}~\bibnamefont {Hradil}},
  \bibinfo {author} {\bibfnamefont {A.}~\bibnamefont {Quattropani}}, \ and\
  \bibinfo {author} {\bibfnamefont {P.}~\bibnamefont {Schwendimann}},\ }\href
  {\doibase 10.1103/PhysRevB.49.8774} {\bibfield  {journal} {\bibinfo
  {journal} {Phys. Rev. B}\ }\textbf {\bibinfo {volume} {49}},\ \bibinfo
  {pages} {8774–8779} (\bibinfo {year} {1994})}\BibitemShut {NoStop}%
\bibitem [{\citenamefont {Abdullah}\ \emph {et~al.}(2013)\citenamefont
  {Abdullah}, \citenamefont {Tang}, \citenamefont {Manolescu},\ and\
  \citenamefont {Gudmundsson}}]{0953-8984-25-46-465302}%
  \BibitemOpen
  \bibfield  {author} {\bibinfo {author} {\bibfnamefont {N.~R.}\ \bibnamefont
  {Abdullah}}, \bibinfo {author} {\bibfnamefont {C.-S.}\ \bibnamefont {Tang}},
  \bibinfo {author} {\bibfnamefont {A.}~\bibnamefont {Manolescu}}, \ and\
  \bibinfo {author} {\bibfnamefont {V.}~\bibnamefont {Gudmundsson}},\ }\href
  {http://stacks.iop.org/0953-8984/25/i=46/a=465302} {\bibfield  {journal}
  {\bibinfo  {journal} {Journal of Physics: Condensed Matter}\ }\textbf
  {\bibinfo {volume} {25}},\ \bibinfo {pages} {465302} (\bibinfo {year}
  {2013})}\BibitemShut {NoStop}%
\bibitem [{\citenamefont {Abdullah}\ \emph {et~al.}(2014)\citenamefont
  {Abdullah}, \citenamefont {Tang}, \citenamefont {Manolescu},\ and\
  \citenamefont {Gudmundsson}}]{Abdullah2014254}%
  \BibitemOpen
  \bibfield  {author} {\bibinfo {author} {\bibfnamefont {N.~R.}\ \bibnamefont
  {Abdullah}}, \bibinfo {author} {\bibfnamefont {C.-S.}\ \bibnamefont {Tang}},
  \bibinfo {author} {\bibfnamefont {A.}~\bibnamefont {Manolescu}}, \ and\
  \bibinfo {author} {\bibfnamefont {V.}~\bibnamefont {Gudmundsson}},\ }\href
  {\doibase 10.1016/j.physe.2014.07.030} {\bibfield  {journal} {\bibinfo
  {journal} {Physica E: Low-dimensional Systems and Nanostructures}\ }\textbf
  {\bibinfo {volume} {64}},\ \bibinfo {pages} {254–262} (\bibinfo {year}
  {2014})}\BibitemShut {NoStop}%
\end{thebibliography}
%

%
%
%----------------------------------------------------------------------------------------
%
\end{document}